\documentclass{emulateapj}

\shorttitle{Star Formation and Environment over Cosmic Time}
\shortauthors{Cooper et al.}

\begin{document}


\title{The DEEP2 Galaxy Redshift Survey: The Role of Galaxy Environment in
  the Cosmic Star--Formation History}

%The Relationship between Star
%  Formation and Galaxy Environment at $\lowercase{z} < 1$}


\author{
Michael C.\ Cooper\altaffilmark{1},
Jeffrey A.\ Newman\altaffilmark{1},
Benjamin J.\ Weiner\altaffilmark{2},
Renbin Yan\altaffilmark{1},
Christopher N.\ A.\ Willmer\altaffilmark{2}, 
Kevin Bundy\altaffilmark{3},
Alison L.\ Coil\altaffilmark{2,7},
Christopher J.\ Conselice\altaffilmark{4},
Marc Davis\altaffilmark{1,5},
S.\ M.\ Faber\altaffilmark{6},
Brian F.\ Gerke\altaffilmark{5},
Puragra Guhathakurta\altaffilmark{6},
David C.\ Koo\altaffilmark{6},
Kai G. Noeske\altaffilmark{6}
}

\altaffiltext{1}{Department of Astronomy, 
University of California at Berkeley, Mail Code
3411, Berkeley, CA 94720 USA; cooper@astro.berkeley.edu, 
jnewman@astro.berkeley.edu, renbin@astro.berkeley.edu, 
marc@astro.berkeley.edu}

\altaffiltext{2}{Steward Observatory, University of Arizona, 
933 N.\ Cherry Avenue, Tucson, AZ 85721 USA; 
bjw@as.arizona.edu, cnaw@as.arizona.edu, acoil@as.arizona.edu}

\altaffiltext{3}{Department of Astronomy \& Astrophysics, University of
  Toronto; bundy@astro.utoronto.ca}

\altaffiltext{4}{University of Nottingham, University Park, 
Nottingham, NG7 2RD, UK; conselice@nottingham.ac.uk}

\altaffiltext{5}{Department of Physics, 
University of California at Berkeley, Mail Code
7300, Berkeley, CA 94720 USA; bgerke@astro.berkeley.edu}

\altaffiltext{6}{UCO/Lick Observatory, UC Santa Cruz, Santa Cruz, CA
95064 USA; faber@ucolick.org, raja@ucolick.org, koo@ucolick.org}

\altaffiltext{7}{Hubble Fellow}

\begin{abstract}

  Using galaxy samples drawn from the Sloan Digital Sky Survey and the
  DEEP2 Galaxy Redshift Survey, we study the relationship between star
  formation and environment at $z \sim 0.1$ and $z \sim 1$. We estimate the
  total star--formation rate (SFR) and specific star--formation rate (sSFR)
  for each galaxy according to the measured [O {\small II}]
  ${\lambda}3727$\AA\ nebular line luminosity, corrected using empirical
  calibrations to match more robust SFR indicators. Echoing previous
  results, we find that in the local Universe star formation depends on
  environment such that galaxies in regions of higher overdensity, on
  average, have lower star--formation rates and longer star--formation
  timescales than their counterparts in lower--density regions. At $z \sim
  1$, we show that the relationship between {\it specific} SFR and
  environment mirrors that found locally. However, we discover that the
  relationship between {\it total} SFR and overdensity at $z \sim 1$ is
  inverted relative to the local relation. This observed evolution in the
  SFR--density relation is driven, in part, by a population of bright, blue
  galaxies in dense environments at $z \sim 1$. This population, which
  lacks a counterpart at $z \sim 0$, is thought to evolve into members of
  the red sequence from $z \sim 1$ to $z \sim 0$. Finally, we conclude that
  environment does not play a dominant role in the cosmic star--formation
  history at $z < 1$: the dependence of the mean galaxy SFR on local galaxy
  density at constant redshift is small compared to the decline in the
  global SFR space density over the last 7 Gyr.

\end{abstract}


%%%%%%%%%%%%%%%%%%%%%%%
%%% 1: Introduction %%%
%%%%%%%%%%%%%%%%%%%%%%%
\section{Introduction}
\label{sec_intro}

The global level (or space density) of star--formation activity has dropped
dramatically from $z \sim 1$ to the present \citep{lilly96, madau96}. While
measurements of the cosmic star--formation history have significantly
improved in precision over the past decade \citep[e.g.,][]{steidel99,
  wilson02, ahopkins04, ahopkins06}, constraining the evolution at $z
\lesssim 1$ to within $\sim \! 50$\%, the cause of this global decline at
late times is still poorly understood.

A wide variety of mechanisms, such as fuel exhaustion via the gradual or
rapid depletion of gas reservoirs or the impact on star formation of a
decline in the galaxy merger rate, have been considered as possible
culprits for the reduction in star--formation activity since $z \sim 1$
\citep[e.g.,][]{lefevre00, hammer05, noeske07b, zheng07}. Many of the
potential causes for the decline in the global star--formation rate should
be closely linked to the environment in which a given galaxy is
found. Physical processes such as ram--pressure stripping or galaxy
harassment, which preferentially occur in regions of higher galaxy density
\citep[e.g.,][]{gunn72, moore96, moore98, hester06}, can remove gas
from galaxies as they fall into rich groups and clusters, leading to a
depletion in star formation via starvation. Heating of intracluster gas due
to cluster mergers \citep[e.g.,][]{mccarthy07} or virial shock heating of
infalling gas in massive dark matter halos \citep[e.g.,][]{birnboim03,
  keres05} could also be responsible for cutting off the supply of cold gas
in high--density environments. Similarly, galaxy groups are the preferred
location for galaxy mergers \citep{cavaliere92} and interactions which may
induce bursts of star formation and/or expulsion of gas
\citep[e.g.,][]{mihos96, cox06, lin07}.

The advent of large spectroscopic galaxy surveys, such as the Sloan Digital
Sky Survey \citep[SDSS,][]{york00} and the 2--degree Field Galaxy Redshift
Survey \citep[2dFGRS,][]{colless01}, has greatly enhanced our ability to
study the connection between galaxies and their environments (determined
from the local overdensity of galaxies). Many galaxy properties ---
including their star--formation rates (SFR) --- have been found to depend
on galaxy environment in the local Universe \citep[e.g.,][]{davis76,
  balogh04a, kauffmann04, christlein05}. For instance, \citet{blanton05}
showed that the typical rest--frame color, luminosity, and morphology of
nearby galaxies is highly correlated with the local galaxy density on $\sim
1\ h^{-1}$ Mpc scales.

As the SDSS and 2dFGRS have revolutionized the study of nearby galaxies,
recent advances in the scope of galaxy surveys at higher redshifts
have permitted some of the first studies of environment able to span a
continuous range of galaxy densities from voids to rich groups and clusters
at $z \sim 1$. Among the current generation of surveys, the DEEP2 Galaxy
Redshift Survey \citep{davis03, faber08} is best suited for studying galaxy
environments at $z \sim 1$, thanks to its relatively large area and
unmatched sample size, number density, and velocity precision.

Studies of galaxies at intermediate redshift have found that many of the
global trends with environment observed locally were in place at $z \sim
1$; using the DEEP2 sample, \citet{cooper06} showed that the color--density
relation was already well--established then, with red galaxies favoring
dense environments relative to their blue counterparts and the bluest
galaxies favoring underdense environments most strongly. Recent results
from COSMOS \citep{scoville07} and the VVDS \citep{lefevre05a} have found
similar trends when looking at both the colors and morphologies of galaxies
at intermediate redshift \citep[e.g.,][]{cucciati06, cassata07}.

The DEEP2 spectroscopy allows measurement of star--formation rates using
the same indicator, [O {\small II}] $\lambda3727$\AA\ luminosity, over the
full primary redshift range of the survey ($0.75 < z < 1.45$). Due to the
high spectral resolution employed $(R \sim 5000)$, this line can be
detected down to relatively low star--formation rates ($\gtrsim 5\ {\rm
  M}_{\sun} {\rm yr}^{-1}$ at $z \sim 1$). Of course, measurements of
luminosities in the ultraviolet are sensitive to dust--extinction
corrections, and the relationship between [O {\small II}] and
star--formation rates should also depend on gas metallicities. However,
using multiwavelength data over wide fields such as the Extended Groth
Strip \citep[EGS,][]{davis07}, [O {\small II}] line luminosities have
recently been calibrated against a variety of star--formation indicators
out to intermediate redshifts, testing the impact of these effects and
improving the robustness of [O {\small II}] star--formation rate estimates
\citep{moustakas06, weiner07}.

In this paper, we utilize galaxy samples drawn from the SDSS and DEEP2
surveys to conduct a detailed study of the relationship between star
formation and environment at both $z \sim 0$ and $z \sim 1$, using as
closely equivalent samples and measurement techniques as possible. Our
principal aim is to investigate the role of environment in the global
decline of the cosmic star--formation rate space density. In \S
\ref{sec_data}, we discuss the data samples employed along with our
measurements of galaxy environments and star--formation rates. Our main
results regarding the relationship between star formation and galaxy
environment are presented in \S \ref{sec_results} and \S
\ref{sec_interpret}. In \S \ref{sec_syst}, we detail possible sources of
contamination. Finally, in \S \ref{sec_disc} and \S \ref{sec_summary}, we
discuss our findings alongside other recent results and summarize our
conclusions. Throughout this paper, we assume a flat $\Lambda$CDM cosmology
with $\Omega_m = 0.3$, $\Omega_{\Lambda} = 0.7$, $w = -1$, and $h=1$ (that
is, a Hubble parameter of $H_0= 100h$ km s$^{-1}$ Mpc$^{-1}$).

% space added to force the heading for Section 2 onto the top of the next
% column, versus being isolated at the extreme bottom of the page.
\vspace*{0.25in}

%%%%%%%%%%%%%%%%%%%%%%
%%% 2: Data Sample %%%
%%%%%%%%%%%%%%%%%%%%%%
\section{The Data Samples}
\label{sec_data}

With spectra for nearly a million galaxies, the Sloan Digital Sky Survey
\citep[SDSS,][]{york00} provides the most expansive picture of the
large--scale structure and local environments of galaxies in the nearby
Universe yet. To study star formation and its relationship with galaxy
density at low redshift $(z \sim 0.1)$, we select a sample of $364,839$
galaxies from the SDSS public data release 4 \citep[DR4,][]{adelman06}, as
contained in the NYU Value--Added Galaxy Catalog
\citep[NYU--VAGC,][]{blanton05b}. We restrict our analysis to galaxies in
the redshift regime $0.05 < z < 0.1$ in an effort to target the nearby
galaxy population while probing a broad range in galaxy luminosity and
simultaneously minimizing aperture effects related to the finite size of
the SDSS fibers. In addition, we limit our sample to SDSS fiber plates for
which the redshift success rate for targets in the main spectroscopic
survey is $80$\% or greater.

\begin{figure*}[t]
\centering
\plotone{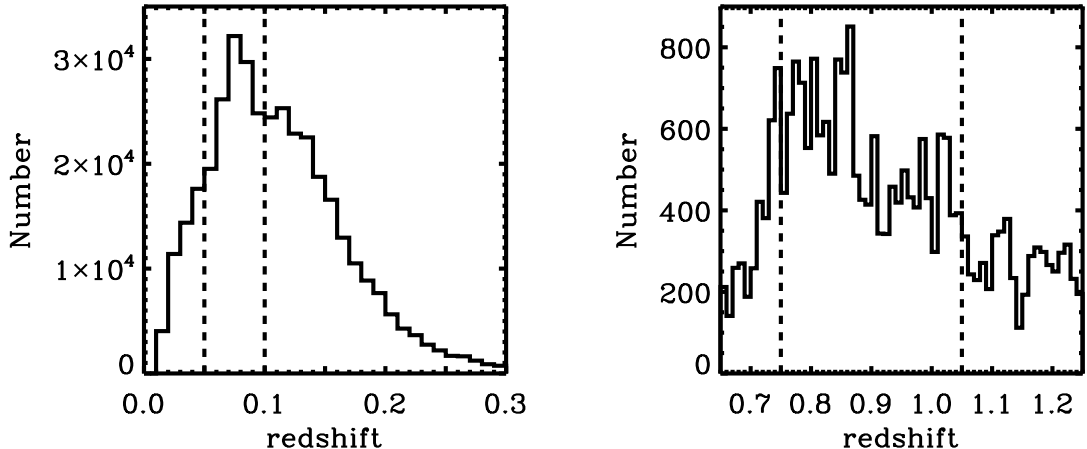}
\caption{(\emph{Left}) The observed redshift distribution for the 374,866
  galaxies drawn from the SDSS within $0.01 < z < 0.3$. (\emph{Right}) The
  observed redshift distribution for the 24,827 DEEP2 galaxies within $0.65
  < z < 1.25$. Both redshift histograms are plotted using a bin size of
  $\Delta z = 0.01$. The dashed vertical lines indicate the redshift ranges
  within each sample used for this paper.}
\label{zdist_fig}
\end{figure*}

In turn, the recently--completed DEEP2 Galaxy Redshift Survey provides the
most detailed census of the Universe at $z \sim 1$ to date. DEEP2 has
targeted $\sim \! 50,000$ galaxies in the redshift range $0 < z < 1.4$ down
to a limiting magnitude of $R_{\rm AB} = 24.1$. Consisting of four widely
separated fields, the survey area covers $\sim \! 3$ square degrees of sky
or roughly 15 times the area of the full moon, with a total of $> \!
25,000$ unique high--precision redshifts from $z = 0.7$ to $z = 1.4$. In
this paper, we utilize a subset consisting of $15,987$ galaxies with
accurate redshifts \citep[quality ${\rm Q} = 3$ or ${\rm Q} = 4$ as defined
by][]{davis07} in the range $0.75 < z < 1.05$ and drawn from all four of
the DEEP2 survey fields. The redshift distributions for the SDSS and DEEP2
galaxy samples used in this paper are plotted in Figure \ref{zdist_fig}.

\subsection{Measurements of Rest--frame Colors and Luminosities}
\label{sec_data1}

For both the SDSS and DEEP2 galaxy samples, we compute rest--frame $U-B$
colors and absolute $B$--band magnitudes ($M_B$) using the {\it kcorrect}
K--correction code (version v4\_1\_2) of \citet[][see also
\citealt{blanton03b}]{blanton07}. The rest--frame quantities for the SDSS
sample are derived from the apparent $ugriz$ data in the SDSS DR4, while
CFHT 12K $BRI$ photometry \citep{coil04b} is used for the DEEP2 sample. All
magnitudes within this paper are on the AB system \citep{oke83} to the
degree to which SDSS magnitudes are AB (as the DEEP2 photometry was
calibrated using SDSS). For conversions between AB and Vega magnitudes, we
refer the reader to Table 1 of \citet{willmer06}.

\begin{figure*}[tb]
\centering
\plotone{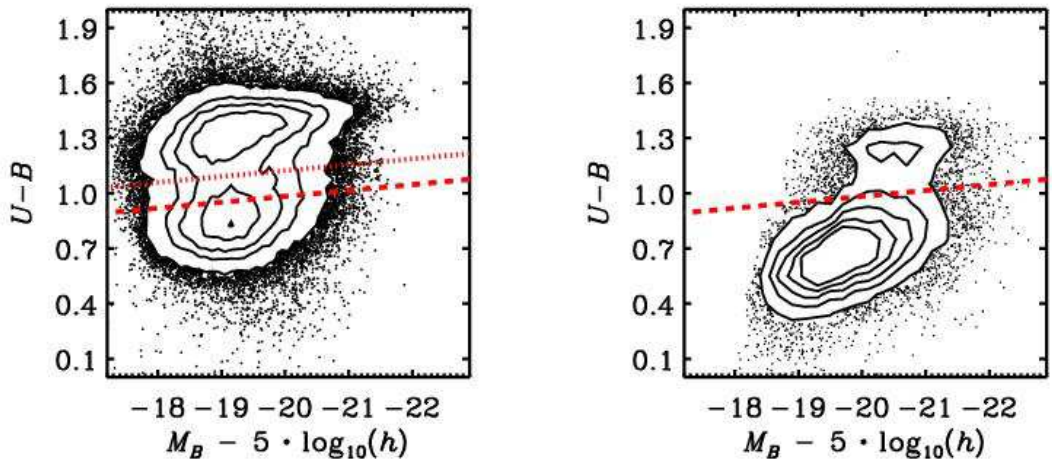}
\caption{The rest--frame $U-B$ versus $M_B$ color--magnitude distributions
  for SDSS galaxies in the redshift range $0.05 < z < 0.1$ (\emph{left})
  and for DEEP2 galaxies in the redshift range $0.75 < z < 1.05$
  (\emph{right}). Due to the large number of galaxies in the each sample,
  we plot contours (rather than individual points) corresponding to 50,
  150, 250, 400, 600, and 1000 galaxies per bin of $\Delta(U-B) = 0.05$ and
  $\Delta M_B = 0.1$ for the SDSS, and corresponding to 50, 100, 150, 200,
  and 250 galaxies per bin of $\Delta(U-B) = 0.1$ and $\Delta M_B = 0.2$
  for DEEP2. Outside of the lowest contour levels, the individual points
  are plotted. The \emph{dashed} red horizontal line shown in both plots
  illustrates the division between the red sequence and the blue cloud used
  at $z \sim 1$, following the relation given in Equation
  \ref{willmer_cut}. The \emph{dotted} red line in the left plot is shifted
  relative to the dashed line by $\Delta(U-B) = 0.14$.}
\label{cmd_fig}
\end{figure*}

The distribution of SDSS and DEEP2 galaxies in $U-B$ versus $M_B$
color--magnitude space is shown in Figure \ref{cmd_fig}. As found by many
previous studies \citep[e.g.,][]{bell04, willmer06}, the galaxy
color--magnitude diagram both at $z \sim 0$ and at $z \sim 1$ exhibits a
clear bimodality in rest--frame color, with a tight red sequence and a more
diffuse ``blue cloud'' of galaxies. We use here the same
magnitude--dependent cut to divide the red sequence and blue cloud at $z
\sim 1$ as employed by \citet{willmer06}; this division in $U-B$ color is
shown in Figure \ref{cmd_fig} as the dashed red line and is given by
\begin{equation}
U - B = -0.032 (M_B + 21.62) + 1.035 .
\label{willmer_cut}
\end{equation}

For the SDSS sample, the red sequence is shifted redward relative to that
of the DEEP2 data by approximately 0.2 magnitudes in $U-B$ color
\citep[cf.\ Fig.\ \ref{cmd_fig} and][]{blanton06a}. This shift is
consistent with the predicted evolution of an old, passively evolving
stellar population, which should redden in $U-B$ color by $\sim \! 0.14$
magnitudes from $z \sim 0.9$ to $z \sim 0.1$ \citep{vandokkum01}. The
dotted red line in Fig.\ \ref{cmd_fig} is simply the DEEP2 division (given
by Equation \ref{willmer_cut}) shifted by $\Delta(U-B) = 0.14$; it provides
a relatively clean divide between the red sequence and blue cloud at $z
\sim 0.1$. We therefore use this shifted line to divide the two for the
SDSS galaxy samples. The blue population, as shown in Figure \ref{cmd_fig},
evolves more with redshift than the red--sequence galaxies, with the blue
cloud being roughly 0.2--0.3 magnitudes redder at $z \sim 0$ relative to $z
\sim 1$. Previous analysis by \citet{blanton06a} found this evolution in
color between the SDSS and DEEP2 to be consistent with the global decline
in the star--formation rate. For a more complete discussion of the
evolution in the color--magnitude distribution of galaxies in SDSS and
DEEP2, we direct the reader to \citet{blanton06a}.

\subsection{Sample Selection}
\label{sec_data2}

As with most deep redshift surveys, the DEEP2 spectroscopic targets span a
broad range in redshift, but were selected according to a fixed
apparent--magnitude limit. The DEEP2 $R_{\rm AB} = 24.1$ magnitude limit
includes different portions of the galaxy population (in rest--frame
color--magnitude space) at different redshifts. To allow tests of how this
selection effect could influence our results, we employ a variety of
subsamples from the full catalog of 15,987 DEEP2 galaxies in the redshift
range $0.75 < z < 1.05$ (which we define to be Sample DEEP2--A).

\begin{figure}[h]
\centering
\plotone{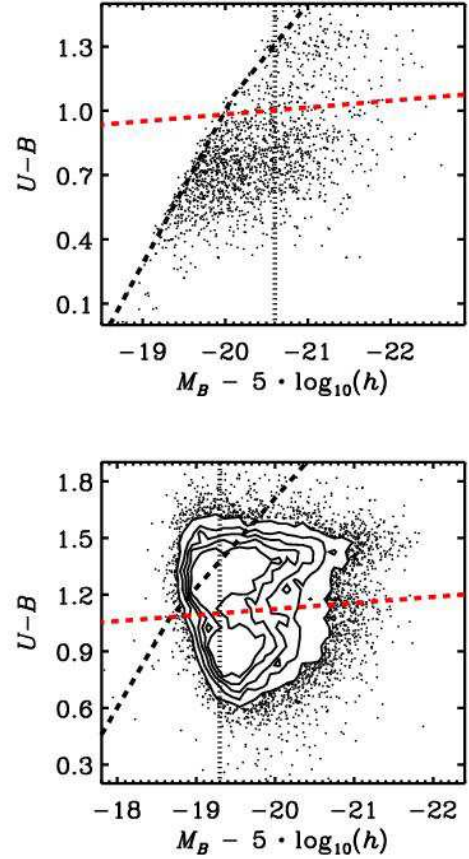}
\caption{(\emph{Top}) The rest--frame color--magnitude diagram for all
  DEEP2 galaxies in the redshift range $1.025 < z < 1.075$. The black
  dotted vertical line defines the color--independent completeness limit of
  the DEEP2 survey at $z = 1.05$, providing the absolute--magnitude limit
  used for Sample DEEP2--C. To this limit, DEEP2 is complete for galaxies
  of all colors at $z < 1.05$. The black dashed line defines the
  completeness limit of the DEEP2 survey as a function of rest--frame color
  at redshift $z = 1.05$ and corresponds to Equation
  \ref{magcut_eqn}. (\emph{Bottom}) The rest--frame color--magnitude
  diagram for all SDSS galaxies in the redshift range $0.095 < z <
  0.105$. The dotted and dashed black vertical lines follow the
  corresponding limits from the top panel, allowing for evolution in
  $M_{B}^{*}$ as given in Equation \ref{magcut_eqn} and Equation
  \ref{magcut2_eqn}. The dashed red line in each plot indicates our
  division between the red sequence and the blue cloud, as given by
  Equation \ref{willmer_cut}. Due to the large number of galaxies in the
  SDSS sample, we plot contours corresponding to 20, 40, 60, 80, and 100
  galaxies per bin of $\Delta(U-B) = 0.05$ and $\Delta M_B = 0.1$.}
\label{sample_select_fig}
\end{figure}

As discussed by \citet{gerke07} and \citet{cooper07a}, it is possible to
produce volume--limited catalogs with a color--dependent
absolute--magnitude cut by defining a region of rest--frame
color--magnitude space that is included by the survey at all redshifts of
interest. For the DEEP2 survey, such a selection cut is illustrated in the
top panel of Figure \ref{sample_select_fig} and given by
\begin{equation}
\begin{array}{l l}
M_{\rm{cut}}(z, U-B) = &
Q(z - z_{\rm lim}) + \\
 & {\rm min}\left\{[a(U-B) + b],\ [c(U-B) + d]\right\}, 
\end{array} 
\label{magcut_eqn}
\end{equation}
where $z_{\rm lim}$ is the limiting redshift beyond which the selected
sample becomes incomplete, $a$, $b$, $c$ and $d$ are constants that are
determined by the limit of the color--magnitude distribution of the sample
with redshift $z > z_{\rm{lim}}$, and $Q$ is a constant that allows for
linear redshift evolution of the typical galaxy absolute magnitude,
$M_{B}^{*}$. For this parameter, we adopt a value of $Q = -1.37$,
determined by \citet{faber07} from a study of the $B$--band galaxy
luminosity function in the COMBO--17 \citep{wolf01}, DEEP1 \citep{vogt05},
and DEEP2 \citep{davis03} surveys. Varying our choice of $Q$ by as much as
$\sim 40\%$ has a negligible effect on our results.

By including this linear $M_{B}^{*}$ evolution in our selection cut, we are
selecting a similar population of galaxies with respect to $M_{B}^{*}$ at
all redshifts. Adopting this approach, with a limiting redshift of $z_{\rm
  lim} = 1.05$, we define a sample of 12,198 galaxies (Sample DEEP2--B)
over the redshift range $0.75 < z < 1.05$ that is volume--limited relative
to $M_{B}^{*}$ and selected according to a color--dependent cut in
$M_{B}$. The values of the constants $a$, $b$, $c$, and $d$ which define
the color--dependent selection are $-1.6$, $-18.65$, $-2.55$, and $-17.7$,
respectively. For complete details of the selection method, we refer the
reader to \citet{gerke07}.

A somewhat simpler selection method is to produce a subsample that is
volume--limited relative to $M_{B}^{*}$ according to a color--independent
cut in absolute magnitude. We create such a sample (Sample DEEP2--C) by
restricting to $0.75 < z < 1.05$ and requiring
\begin{equation}
M_B(z) \le -20.6 + Q (z - z_{\rm lim}), 
\label{magcut2_eqn}
\end{equation}
where $M_B = -20.6$ is the absolute magnitude to which DEEP2 is complete
along both the red sequence and the blue cloud at $z_{\rm lim} = 1.05$
(cf.\ the top panel of Figure \ref{sample_select_fig}). A brief summary of
all galaxy samples utilized in this paper is provided in Table
\ref{sample_descript_tab}.

\begin{deluxetable*}{l l l l l l}
\tablewidth{0pt}
\tablecolumns{6}
\tablecaption{\label{sample_descript_tab} Descriptions of the SDSS and
  DEEP2 Galaxy Samples} 
\tablehead{ Sample & $N_{\rm galaxies}$ & $N_{\rm edge-cut}$ & $N_{\rm
    SFR}$ & $z$ range & Brief Description }
\startdata
DEEP2--A & 15,987 & 12,240 & 11,875 & $0.75 < z < 1.05$ & 
\parbox[l]{2.75in}{all galaxies after boundary cut} \\
 &  &  &  &  & \\
DEEP2--B & 12,198 & 9,346 & 9,067 & $0.75 < z < 1.05$ & 
\parbox[l]{2.75in}{color--dependent limit, with limit held 
constant relative to $M_{B}^{*}(z)$} \\
 &  &  &  &  & \\
DEEP2--C & 4,387 & 3,349 & 3,178 & $0.75 < z < 1.05$ & 
\parbox[l]{2.75in}{color--independent limit, with limit held constant
  relative to $M_{B}^{*}(z)$ and set as $M_{B} = -20.6$ at $z =
  1.05$}\\
&  &  &  &  & \\
SDSS--A & 132,367 & 122,577 & 120,636 & $0.05 < z < 0.1$ &
\parbox[l]{2.75in}{all galaxies after boundary cut} \\
 &  &  &  &  & \\
SDSS--B & 61,413 & 57,051 & 56,118 & $0.05 < z < 0.1$ & 
\parbox[l]{2.75in}{color--independent limit, with limit held constant
  relative to $M_{B}^{*}(z)$ and set as $M_{B} = -20.6$ at $z =
  1.05$}\\
&  &  &  &  & \\
SDSS--C & 42,991 & 39,978 & 39,254 & $0.05 < z < 0.1$ &
\parbox[l]{2.75in}{color--independent limit, with limit held constant
  relative to $M_{B}^{*}(z)$ and set as $M_{B} = -20.6$ at $z =
  1.05$; matched to DEEP2 red fraction}\\

\enddata
\tablecomments{We list each galaxy sample employed in the analysis,
  detailing the selection cut used to define the sample as well as the
  redshift range covered and the number of galaxies included before
  $(N_{\rm galaxies})$ and after $(N_{\rm edge-cut})$ removing those within
  $1\ h^{-1}$ comoving Mpc of a survey edge. The number of galaxies with an
  accurate SFR measurement and away from a survey edge is given by $N_{\rm
    SFR}$.}
%\label{sample_descript_tab}
\end{deluxetable*}

To facilitate the comparison of trends with local environment at $z \sim
0.1$ to those at $z \sim 1$, we select subsamples drawn from the SDSS which
mimic the DEEP2 survey subsamples detailed above. While both the SDSS and
DEEP2 spectroscopic targets are selected according to an
apparent--magnitude limit in a red optical band ($r \le 17.77$ and $R \le
24.1$, respectively), this band falls in a very different part of the
spectrum in the rest--frame at the two redshift ranges probed. At $z \sim
0.1$ the center of the SDSS $r$ passband corresponds to a rest--frame
wavelength of $5605$\AA, whereas at $z \sim 1$ the CFHT $R$ passband
employed by DEEP2 samples a portion of rest--frame wavelength space
centered on roughly $3300$\AA, well into the ultraviolet. As a result, the
DEEP2 sample is biased towards blue (in rest--frame $U-B$ color) galaxies
relative to the SDSS data set; in fact, DEEP2 probes to much fainter
luminosities on the blue cloud (relative to $L_{B}^{*}$) than the SDSS
sample, as shown in Figure \ref{sample_select_fig}. For this reason, we are
unable to define an SDSS subsample that totally matches the DEEP2--B
sample.

However, we can define two subsamples drawn from the full SDSS data set of
132,367 galaxies at $0.05 < z < 0.1$ (Sample SDSS--A) which complement the
DEEP2--C sample. First, we select an SDSS subsample (Sample SDSS--B) that
adheres to the same selection limit as the DEEP2--C galaxy sample. That is,
we define an SDSS sample that is volume--limited relative to $M_{B}^{*}$,
according to the color--independent cut in absolute magnitude given in
Equation \ref{magcut2_eqn}.

Within both the SDSS and DEEP2 galaxy catalogs, the bimodality of galaxy
colors in rest--frame $U-B$ color is clearly visible (cf.\ Fig.\
\ref{cmd_fig} and Fig.\ \ref{sample_select_fig}). To quantify the
composition of the SDSS and DEEP2 data sets in terms of red and blue
galaxies, we compute the fraction of galaxies on the red sequence in each
survey sample using the color divisions defined above (cf. Equation
\ref{willmer_cut}, offset by 0.14 magnitudes for the SDSS as described in
\S \ref{sec_data1}).

Studies of the galaxy luminosity function at $z < 1$ have shown that the
number density of galaxies on the red sequence has increased over the last
7 Gyr, yielding an increase in the red galaxy luminosity density of
$\gtrsim$ a factor of 2 \citep{bell04, faber07}. Meanwhile, the luminosity
and number density of galaxies on the blue cloud has remained roughly
constant (especially relative to that of the red sequence) over the same
timespan.\footnote{There is some debate within the community regarding the
  evolution in the number density of blue galaxies at intermediate and low
  redshift. Parallel studies of the luminosity and stellar mass functions
  at $z < 1$ have found significant evolution in the number density of
  bright (massive), blue galaxies \citep[e.g.,][]{bundy06, zucca06}.} Thus,
the relative fractions of red and blue galaxies in magnitude--limited
samples will vary with redshift. Using the divisions between red and blue
galaxies defined above, the fraction of galaxies which are on the red
sequence in samples DEEP2--C and SDSS--B is $0.39$ and $0.56$,
respectively. Because red--sequence galaxies are forming few stars, we
might expect the overall average SFR in galaxies in the SDSS to be lower
simply due to this greater fraction of quiescent galaxies, rather than
through a modulation of the rate in star--forming objects.

\begin{figure}[h]
\centering
\plotone{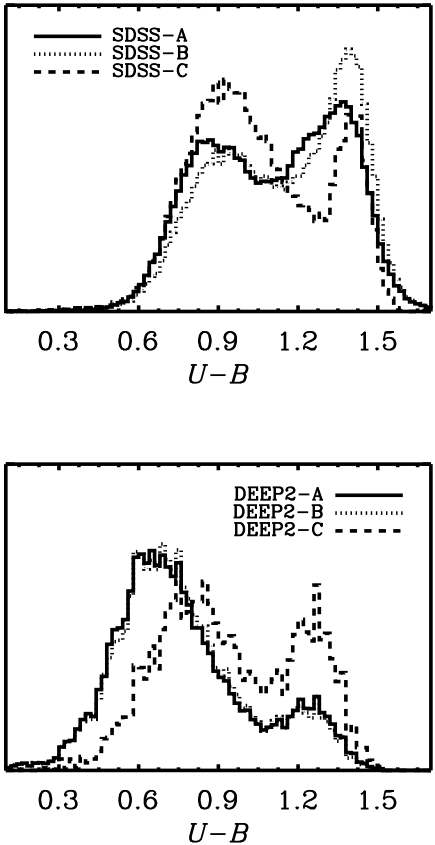}
\caption{The relative distribution of rest--frame $U-B$ colors for
  the galaxy subsamples listed in Table \ref{sample_descript_tab}. By
  randomly excluding red galaxies from the SDSS sample, we are able to define a
  subsample (SDSS--C) with a red--galaxy fraction comparable to that of the
  DEEP2--C sample. The plotted histograms have been normalized to have
  equal area.}
\label{color_dist_fig}
\end{figure}

\begin{figure}[h]
\centering
\plotone{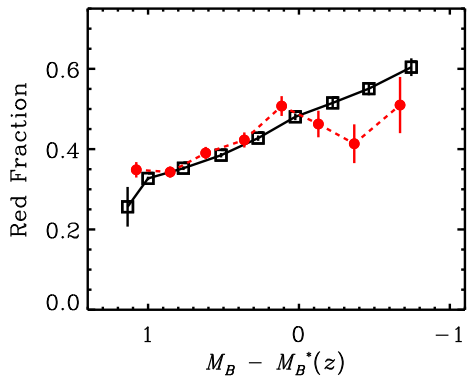}
\caption{The fraction of red galaxies as a function of $B$--band absolute
  magnitude [relative to $M_{B}^{*}(z)$] for the SDSS--C (\emph{open black
    squares}) and DEEP2--C (\emph{filled red circles}) samples (i.e., the
  galaxy samples constructed to have color--independent $M_B$ cuts and
  equivalent aggregate red--galaxy fractions). The red galaxy population is
  selected in each survey sample using the color divisions defined by
  Equation \ref{willmer_cut}, offset by 0.14 magnitudes for the SDSS, with
  the error in the red fraction given by binomial statistics. We assume an
  evolution of 1.37 magnitudes per unit redshift $(Q = -1.37)$ in
  $M_{B}^{*}(z)$, with $M_{B}^{*}(z=0.9) = -21.48$ from
  \citet{willmer06}. }
\label{mag_dist_fig}
\end{figure}

To select a sample from the SDSS that is more analogous to the DEEP2--C
sample (i.e., yielding an equivalent red fraction down to the common
magnitude limit), we randomly throw out red galaxies from SDSS--B. The
resulting sample (SDSS--C) contains 42,991 galaxies with a distribution of
rest--frame colors comparable to that of DEEP2--B, as shown in Figure
\ref{color_dist_fig}, and a red fraction of $0.39$. We have not required
the SDSS galaxies to follow the same absolute--magnitude distribution as
the DEEP2--C sample. However, the dependence of the fraction of red (or
blue) galaxies on $M_B - M_B^{*}$ in the SDSS--C and DEEP2--C samples are
very similar (as shown in Figure \ref{mag_dist_fig}), with blue galaxies
dominating at faint luminosities and with blue and red populations each
comprising roughly half of the population at the bright end of the $M_B$
distribution. A summary of both the DEEP2 and SDSS galaxy samples is
provided in Table \ref{sample_descript_tab}.

\subsection{Measurements of Local Galaxy Environment}
\label{sec_data3}

For the purposes of this paper, we consider the ``environment'' of a galaxy
to be defined by the local mass overdensity, measured using the local
overdensity of galaxies as a proxy; over quasi--linear regimes, these
should differ by a factor of the galaxy bias \citep{kaiser87}. We estimate
this overdensity for both the SDSS and DEEP2 using measurements of the
projected $3^{\rm rd}$--nearest--neighbor surface density $(\Sigma_3)$
about each galaxy, where the surface density depends on the projected
distance to the $3^{\rm rd}$--nearest neighbor, $D_{p,3}$, as $\Sigma_3 = 3
/ (\pi D_{p,3}^2)$. In computing $\Sigma_3$, a velocity window of $\pm
1000\ {\rm km}/{\rm s}$ is employed to exclude foreground and background
galaxies along the line--of--sight. Tests by \citet{cooper05} found this
environment estimator to be a robust indicator of local galaxy density for
the DEEP2 survey.

To correct for the redshift dependence of the sampling rate of both the
SDSS and the DEEP2 surveys, each surface density value is divided by the
median $\Sigma_3$ of galaxies at that redshift within a window of $\Delta z
= 0.02$ and $\Delta z = 0.04$ for the SDSS and DEEP2, respectively; this
converts the $\Sigma_3$ values into measures of overdensity relative to the
median density (given by the notation $1 + \delta_3$ here) and effectively
accounts for redshift variations in the selection rate \citep{cooper05}. In
computing the local environment for galaxies in our targeted redshift
ranges ($0.05 < z < 0.1$ for the SDSS and $0.75 < z < 1.05$ for DEEP2), we
included sources at lower and higher redshifts as tracers of the galaxy
distribution to avoid edge effects due to redshift limits; similarly, the
smoothing windows for calculations of median $\Sigma_3$ include tracers
outside the sample $z$ limits.

\begin{figure}[h!]
\centering
\plotone{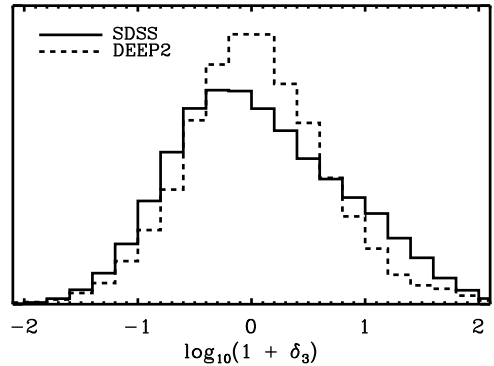}
\caption{The distribution of the logarithm of the local overdensities,
  $\log_{10}{(1 + \delta_3)}$,for the DEEP2 and SDSS samples. We plot the
  environment distributions for the 12,240 DEEP2 galaxies with $0.75 < z <
  1.05$ and more than $1\ h^{-1}$ comoving Mpc from a survey edge (solid
  line) along with that for the 132,367 SDSS galaxies with $0.05 < z < 0.1$
  and more than $1\ h^{-1}$ comoving Mpc from a survey edge (dashed
  line). The overdensity, (1 + $\delta_3$), is a dimensionless quantity,
  computed as described in \S \ref{sec_data3}. Here, we scale the DEEP2 and
  SDSS histograms so that their integrals are equal.}
\label{delta3_fig}
\end{figure}

Finally, to minimize the effects of edges and holes in the SDSS and DEEP2
survey geometries, we exclude all galaxies from our SDSS and DEEP2 samples
within $1\ h^{-1}$ Mpc (comoving) of a survey boundary, reducing our sample
sizes to the numbers given in Table \ref{sample_descript_tab}. In Figure
\ref{delta3_fig}, we plot the distribution of overdensities for the SDSS--A
and DEEP2--A samples, after these edge cuts. For complete details regarding
the computation of the local environment measures, we direct the reader to
\citet{cooper06}.

The overdensity distributions for the SDSS and DEEP2, as shown in Fig.\
\ref{delta3_fig}, differ for several reasons. The first is simply the
nonlinear growth of large--scale structure over time: a dense region on
nonlinear scales will be denser at $z \sim 0$ than at $z \sim 1$, while a
void will be less dense today than in the past.  A second reason is that
the average bias of the overall SDSS sample used as a tracer of density is
higher than the overall DEEP2 sample \citep{zehavi02, zehavi05, coil04a,
  coil06}; this will cause density contrasts measured with galaxies to be
exaggerated in the SDSS compared to DEEP2.

A third cause for the differences in these distributions is that we are
using the projected $3^{\rm rd}$--nearest--neighbor distance, $D_{p,3}$ to
measure overdensity in both samples; but because the number density of the
SDSS sample used to trace environment is higher than in DEEP2, the typical
$D_{p,3}$ for the SDSS is smaller ($\sim 1\ h^{-1}$ comoving Mpc) than for
a DEEP2 galaxy ($\sim 1.8\ h^{-1}$ comoving Mpc). Hence, in the SDSS, we
are measuring overdensities on somewhat smaller, more highly nonlinear
scales. However, \citet{blanton06b} found that in the SDSS, environments
measured on scales from 0.2 to 6 $h^{-1}$ Mpc yield equivalent results; we
therefore do not expect this to be a major issue.

All of these effects operate in the same sense, exaggerating the density
contrasts measured in the SDSS. However, none of them should change the
rank ordering of overdensities. In this paper, we focus on changes in the
general relationships between environment and galaxy properties (namely,
star--formation activity) between $z \sim 0$ and $z \sim 1$; this requires
only that we have an accurate measure of relative environment at each
redshift. The $(1 + \delta_3)$ values provide such a tracer of local galaxy
density at both epochs.

\subsection{Measurements of Star--Formation Rates}
\label{sec_data4}

We estimate the global star--formation rates for galaxies in both the SDSS
and DEEP2 samples using measured [O {\small II}] ${\lambda}3727$\AA\
nebular line luminosities, corrected using the empirical calibration of
\citet{moustakas06}. Although this calibration was developed by tuning [O
{\small II}]--derived star--formation rates (for dust extinction and
metallicity) to match those based on extinction--corrected H$\alpha$ and
far--infrared emission using multiwavelength observations of nearby
galaxies, the results have been tested and proven effective at intermediate
redshifts $(0.7 < z < 1.4)$ \citep{moustakas06}. As will be shown in \S
\ref{sec_syst3}, however, the results presented in this paper are not
sensitive to the particular calibration employed. Using star--formation
rates derived with $1/2 \times$ or $2 \times$ the luminosity--dependence
given by \citet{moustakas06}, accounting for a wide range in possible dust
and metallicity effects, using the calibrations of \citet{kennicutt98} and
\citet{weiner07}, or using the SFR estimates of \citet{tremonti04}, we find
that our results regarding the relationships between star formation, color,
and environment at $z \sim 0.1$ and $z \sim 1$ remain unchanged. For both
the SDSS and DEEP2 galaxy samples, all estimated star--formation rates are
given in units of $h^{-2} {\rm M}_{\sun} {\rm yr}^{-2}$. As noted in Table
\ref{sample_descript_tab}, a small number of objects in both the SDSS and
DEEP2 were rejected from the galaxy samples due to large uncertainties in
their measured [O {\small II}] line fluxes ($\sigma_{\rm [O II]} > 500
\times 10^{-19}\ {\rm erg} {\rm s}^{-1} {\rm cm}^2$).

For the DEEP2 sample, we measure [O {\small II}] equivalent widths using a
nonlinear least--squares fit to the observed emission lines in the
DEEP2/DEIMOS spectra, using a model given by two Gaussians of the same
width $(\sigma)$, centered at the known rest--frame wavelengths of the two
components of the doublet. The continuum level is estimated from the
biweight of the continuum in two windows, 15--60\AA\ away from the emission
line in the rest--frame. The observed $R-I$ color and $I$ magnitude for
each galaxy is used to estimate its continuum luminosity at 3727\AA\ via
the K--correction procedure of \citet{willmer06}. Combined with the
measured [O {\small II}] equivalent width, this yields a flux--calibrated
line luminosity. The line luminosities are then transformed into
star--formation rates using a correction factor based upon the galaxy's
$B$--band absolute magnitude, as given by a linear interpolation of the
values in Table 2 of \citet{moustakas06}; we test the impact of using other
conversions in \S \ref{sec_syst3}. For complete details regarding the
computation of [O {\small II}] emission--line luminosities in DEEP2, see
\citet{weiner07}.

We estimate the [O {\small II}] luminosities for SDSS galaxies by measuring
the total observed line flux in the spectrum. This gives the [O {\small
  II}] luminosity integrated over the area covered by the SDSS fiber, which
might be an underestimate of a galaxy's true total [O {\small II}]
luminosity, given the limited angular size of the SDSS
fibers. Alternatively, we can estimate the true total [O {\small II}]
luminosity by combining a measurement of the [O {\small II}] equivalent
width with the K--corrected $u$--band absolute magnitude (i.e., assuming
that the ratio of [O {\small II}] flux to $u$ flux is uniform across the
entire galaxy). This yields a significantly less precise measure of the [O
{\small II}] luminosity, due to the high noise level in SDSS $u$--band
photometry. Nevertheless, if we were to adopt these noisier SFR estimates,
none of our conclusions would be changed.

We measure the [O {\small II}] line flux for SDSS galaxies in a 22\AA\
window around the line after removing all stellar continuum features from
the spectra. The stellar continua around 3727\AA\ are very bumpy and would
introduce systematic [O {\small II}] flux offsets if not accurately
subtracted. The subtraction procedure used is described in \citet{yan06};
we summarize here. After subtracting off the continuum of the spectrum
smoothed over a broad window, the stellar continuum is fit to a linear
combination of two stellar population templates produced with Bruzual \&
Charlot models \citep{bc03}, again with their broad continuum components
subtracted. One template is the spectrum of a 7--Gyr--old simple stellar
population, while the other corresponds to a 0.3--Gyr--old starburst of
duration 0.1 Gyr (i.e., a starburst commencing 0.4 Gyr in the past). This
combination of templates, each constructed with solar metallicity, has
proven adequate to accurately describe the wiggles in the continuum near [O
{\small II}] for most galaxies in the SDSS. With the stellar continuum
features removed, we measure the line flux in the remaining, emission--line
only spectrum. The uncertainties in this continuum subtraction have been
propagated into our error estimates for [O {\small II}] fluxes.

\subsection{Measurements of Stellar Masses}
\label{sec_data5}

Stellar masses for the SDSS galaxies were determined using the {\it
  kcorrect} K--correction code of \citet{blanton03b}. The template SEDs
employed by {\it kcorrect} are based on those of \citet{bc03}; the
best--fit SED given the observed $ugriz$ photometry and spectroscopic
redshift can be used directly to estimate the stellar mass--to--light ratio
$({\rm M}_{*}/L)$, assuming a \citet{charbrier03} initial mass function.

For a portion of the DEEP2 galaxy catalog, stellar masses may be calculated
using WIRC/Palomar $J$-- and $K_s$--band photometry in conjunction with the
DEEP2 $BRI$ data \citep{bundy06}. The observed ($BRIJK_s$) SED of each
$K_s$--detected galaxy is compared to a grid of 13440 synthetic SEDs from
\citet{bc03}, which span a range of star--formation histories, ages,
metallicities, and dust content, and use a \citet{charbrier03} initial mass
function \citep{bundy05}. From fits to the grid of models, a stellar--mass
probability distribution is obtained after scaling each model's ${\rm
  M}_{*}/L_K$ ratio to the total $K_s$ magnitude and marginalizing over the
grid. The median of this distribution is taken as the stellar mass estimate
\citep{bundy06}.

The $K_s$--band photometry, however, does not cover the entire area of the
DEEP2 survey, and often faint blue galaxies at the high--$z$ end of the
DEEP2 redshift range are not detected in $K_s$. Because of these two
effects, the stellar masses of \citet{bundy06} have been used to calibrate
stellar mass estimates for the full DEEP2 sample that are based on
combining rest--frame $M_B$ and $B-V$ derived from the DEEP2 data
\citep{lin07} into the expressions of \citet{bell03}, which use a ``diet
Salpeter'' IMF and are valid at $z = 0$. We empirically correct these
stellar mass estimates to the \citet{bundy06} measurements by accounting
for a mild color and redshift dependence \citep{lin07}; where they overlap,
the two stellar masses have an RMS difference of approximately 0.3 dex
after this recalibration.

\begin{figure*}[tb]
\centering
\plotone{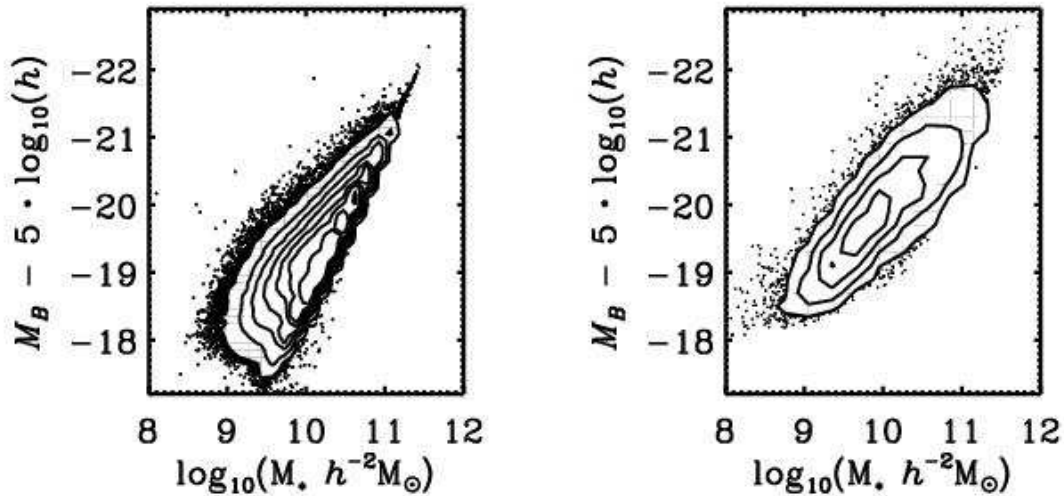}
\caption{The relationship between stellar mass and absolute $B$--band
  magnitude for galaxies in the SDSS--A (\emph{left}) and DEEP2--A
  (\emph{right}) samples. There is a strong correlation between $M_B$ and
  ${\rm M}_{*}$ in both galaxy samples. The contours correspond to 50, 150,
  300, 500, 750, 1000, and 1500 galaxies per bin of $\Delta
  (\log_{10}{M_{*}}) = 0.15$ and $\Delta M_B = 0.1$ (\emph{left}) and 50,
  150, 300, 500, and 1000 galaxies per bin of $\Delta (\log_{10}{M_{*}}) =
  0.3$ and $\Delta M_B = 0.2$ (\emph{right}).}
\label{mass_mag_fig}
\end{figure*}

\begin{figure*}[tb]
\centering
\plotone{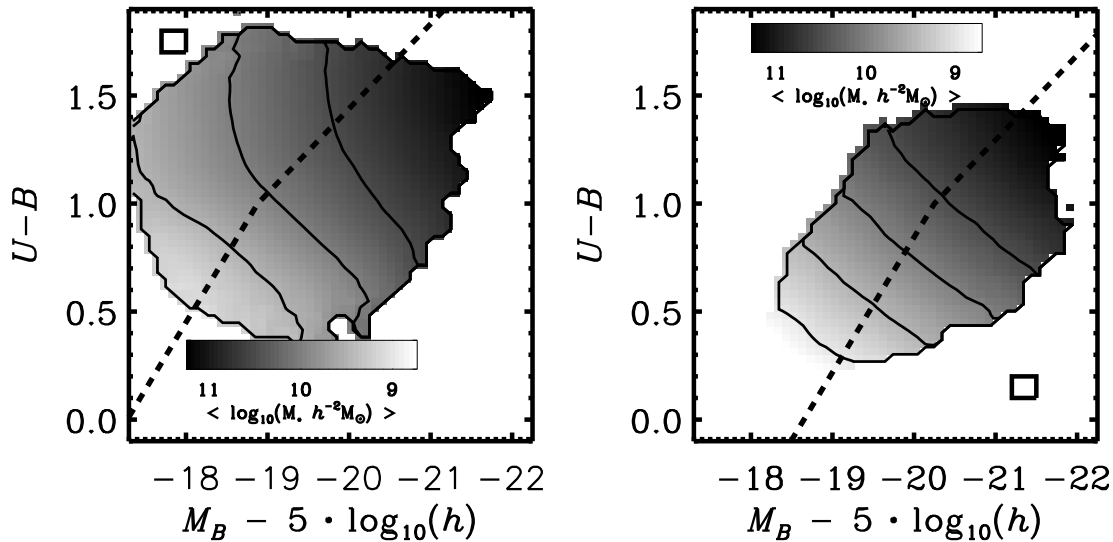}
\caption{The mean stellar mass as a function of galaxy color, $U-B$, and
  absolute magnitude, $M_B$, for galaxies in the SDSS--A (\emph{left}) and
  DEEP2--A (\emph{right}) samples. The means are computed in a sliding box,
  illustrated in the corner of each plot, with width $\Delta M_B = 0.2$ and
  height $\Delta (U-B) = 0.1$. Darker areas in the image correspond to
  regions of higher average stellar mass in color--magnitude space with the
  scale given by the corresponding inset color bar. The contours in both
  plots correspond to levels of $<\log_{10}({\rm M_{*}})> = $9, 9.5, 10,
  and 10.5. At regions where the sliding box includes fewer than 20
  galaxies, the mean stellar mass is not displayed. The dashed lines in
  each plot show the color--dependent, absolute--magnitude selection cut
  (cf.\ Equation \ref{magcut_eqn}) used in defining sample DEEP2--B.}
\label{mean_mass_cmd}
\end{figure*}

We note that while rest--frame $B$--band emission is more sensitive to the
presence of young stars than redder bands, there is still a strong
correlation between stellar mass and absolute $B$--band magnitude in both
the SDSS and DEEP2 samples. As shown in Figure \ref{mass_mag_fig} and
Figure \ref{mean_mass_cmd}, the RMS difference between ${\rm M}_{*}$ and
$M_B$ is roughly 0.5 dex; these differences are strongly correlated with
rest--frame galaxy color.

By combining the measurements of total SFR and stellar mass estimates
described above, we can compute the specific star--formation rate (sSFR)
for each galaxy in the SDSS and DEEP2 samples. The sSFR describes the
fractional rate of stellar mass growth (${\rm sSFR} = {\rm SFR}/ {\rm
  M}_{*}$) in a galaxy due to ongoing star formation. The sSFR has units of
inverse time; for this reason, galaxies with low specific star--formation
rates are said to have long star--formation timescales and vice versa.

%%%%%%%%%%%%%%%%%%
%%% 3: Results %%%
%%%%%%%%%%%%%%%%%%
\section{Results}
\label{sec_results}

Because it is expected that the local environment of a galaxy should
influence its properties (such as star--formation rate), it is common to
study those properties as a function of environment. Since our tracers of
environment are generally sparse, however, measurements of galaxy densities
are generally significantly more uncertain than measures of most other
properties such as color, luminosity, or even SFR. Therefore, binning
galaxies according to local overdensity introduces a significant
correlation between neighboring environment bins, which can smear out any
underlying trends. In this paper, we study both the dependence of mean
environment on galaxy properties and vice versa: the former minimizes
covariance, while the latter eases comparison to other studies. Throughout
\S \ref{sec_results}, we show results for the SDSS--A and DEEP2--A samples,
which probe the greatest range in luminosity and have the largest sample
sizes. However, the qualitative relationships between environment and star
formation for each of the galaxy subsamples in Table
\ref{sample_descript_tab} are consistent with each other, with the
normalization and strength of the trends varying amongst them (cf.\ Table
\ref{ssfr_fits_tab} and Table \ref{sfr_fits_tab}).  We present our
principal results in this section; these results will be interpreted in
$\S$ \ref{sec_interpret} and the remainder of the paper.

\subsection{The sSFR--density relation at $z \sim 0.1$ and $z \sim 1$}
\label{sec_results_ssfr}

\begin{figure*}[tb]
\centering
\plotone{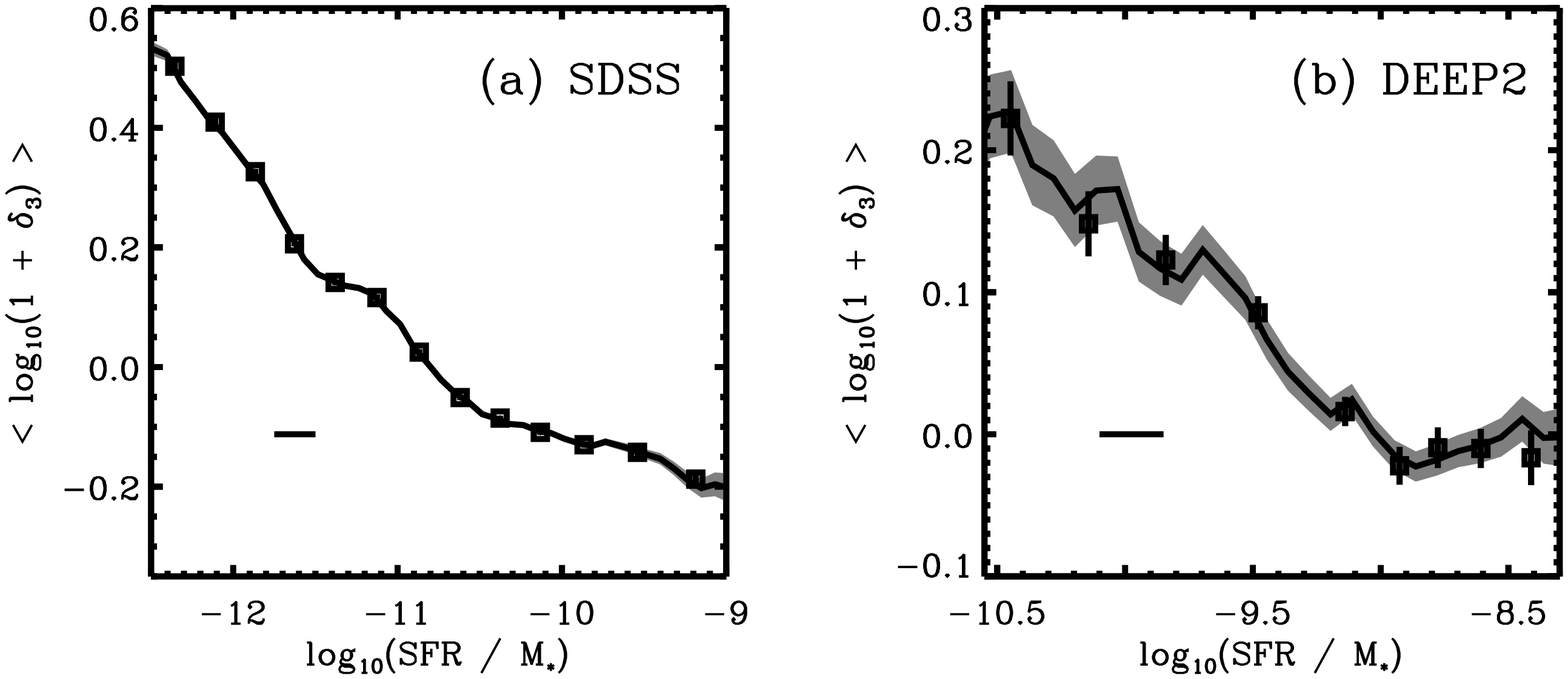}
\caption{The relationship between mean environment and sSFR at $z \sim 0.1$
  in the SDSS (\emph{left}) and at $z \sim 1$ in DEEP2 (\emph{right}). We
  plot the mean and the error in the mean of the logarithm of the local
  galaxy overdensity in discrete bins of sSFR (square points). The solid
  black lines show the mean dependence of environment on sSFR, where the
  means were computed using sliding boxes with widths given by the black
  dashes in the plot. The accompanying grey regions correspond to the
  sliding 1$\sigma$ uncertainties in the means.}
\label{environ_ssfr_fig}
\end{figure*}

The connection between specific star--formation rate and environment at $z
\sim 0$ has been explored by a number of previous studies utilizing
samples drawn from the SDSS or other catalogs of nearby galaxies
\citep[e.g.,][]{lewis02, gomez03, kauffmann04}. In agreement with these
earlier analyses, we find that galaxies with lower specific star--formation
rates, on average, favor regions of higher galaxy density at $z \sim 0.1$
(cf.\ Figure \ref{environ_ssfr_fig}a). 

Using the DEEP2 data set to study galaxy properties at $z \sim 1$, we find
that the dependence of mean environment on sSFR, as found in the SDSS--A
sample, is echoed in the DEEP2--A sample. As shown in Figure
\ref{environ_ssfr_fig}, galaxies with longer star--formation timescales
(i.e., lower specific star--formation rates) favor regions of higher galaxy
density at both $z \sim 0.1$ and $z \sim 1$.

\begin{figure*}[tb]
\centering
\plotone{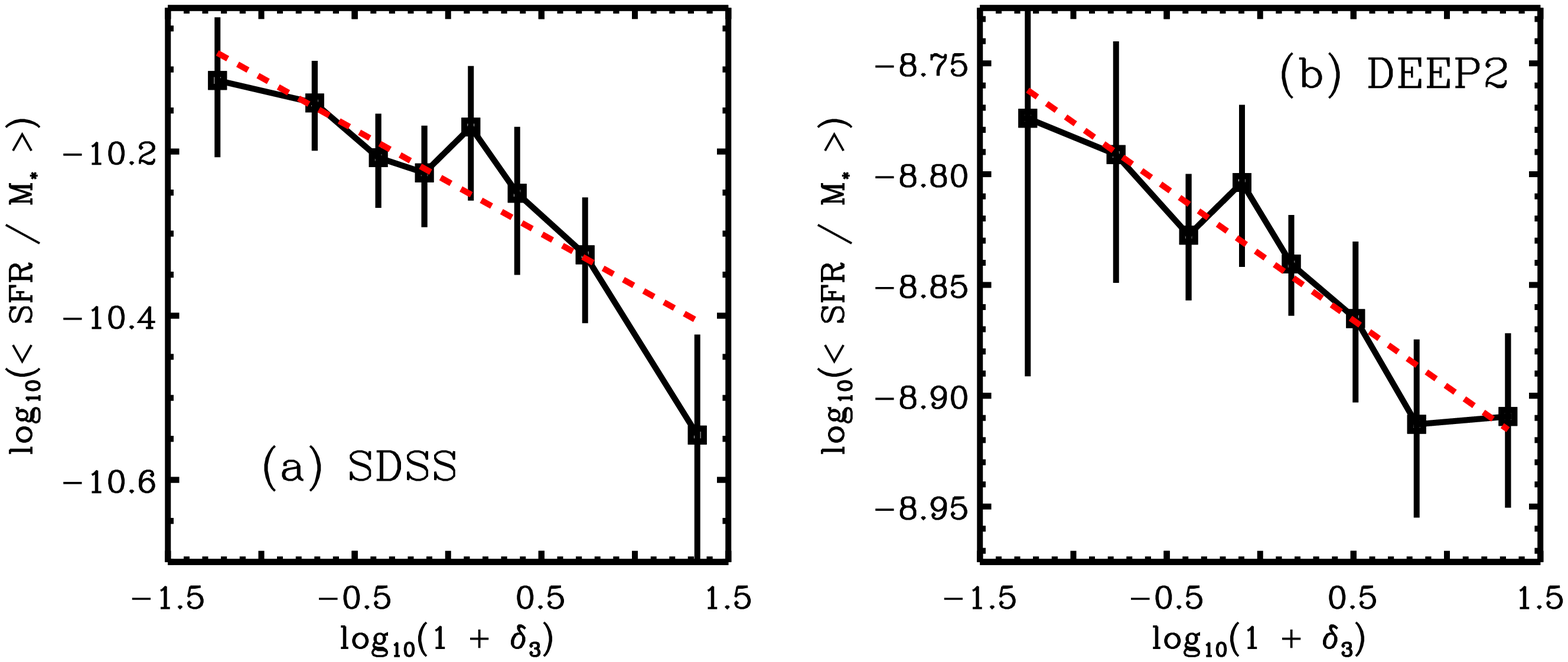}
\caption{The dependence of mean sSFR on environment at $z \sim 0.1$ in the
  SDSS (\emph{left}) and at $z \sim 1$ in DEEP2 (\emph{right}). We plot the
  logarithm of the mean and of the error in the mean of the sSFR in
  discrete bins of galaxy overdensity within the SDSS--A and DEEP2--A
  samples. The dashed red line in each plot shows a least--squares
  linear--regression fit to the data points, with coefficients of the fits
  given in Table \ref{ssfr_fits_tab}.}
\label{ssfr_environ_fig}
\end{figure*}

The same general trend is found when we examine the connection between sSFR
and environment from the opposite perspective. Figure
\ref{ssfr_environ_fig} shows the dependence of the mean galaxy sSFR on
local overdensity in the SDSS--A and DEEP2--A samples. We find that, at $z
\sim 0.1$ and at $z \sim 1$, galaxies residing in regions of higher density
generally have lower specific star--formation rates. Moving to
higher--density environments at $z \sim 0.1$ and at $z \sim 1$, the average
specific star--formation rate declines monotonically such that members of
clusters and massive groups, as a population, exhibit the longest
star--formation timescales (or the lowest fractional rate of stellar mass
growth).

All of the SDSS and DEEP2 samples described in \S \ref{sec_data2} exhibit a
highly significant anticorrelation between specific star--formation rate
and galaxy environment. Table \ref{ssfr_fits_tab} lists the coefficients
from linear--regression fits to the dependence of mean sSFR on overdensity
in each galaxy sample. The fits to the SDSS--A and DEEP2--A samples are
shown in Figure \ref{ssfr_environ_fig} as the dashed red lines. The slopes
of the trends between mean sSFR and overdensity for each of the SDSS and
DEEP2 subsamples agree within the uncertainties. Variations in
normalization between subsamples are associated with differences in sample
selection and galaxy evolution at $z < 1$; these effects are examined in
more detail in \S \ref{sec_disc}.

\begin{deluxetable}{l l l l l}
\tablewidth{0pt}
\tablecolumns{5}
\tablecaption{\label{ssfr_fits_tab} Fits to sSFR--density Relation}
\tablehead{ Sample & $a_0$ & $a_1$ & $\sigma_{a_0}$ & $\sigma_{a_1}$}
\startdata
DEEP2--A & -8.84 & -0.060 & 0.013 & 0.020 \\
DEEP2--B & -8.81 & -0.067 & 0.011 & 0.017 \\
DEEP2--C & -9.05 & -0.067 & 0.007 & 0.011 \\
SDSS--A & -10.24 & -0.128 & 0.024 & 0.036 \\
SDSS--B & -10.31 & -0.124 & 0.023 & 0.033 \\
SDSS--C & -10.19 & -0.087 & 0.024 & 0.035 \\
\enddata
\tablecomments{We list the coefficients and 1$\sigma$ uncertainties for the
  parameters of the linear--regression fits to the sSFR--density relation
  given by $\log_{10}(<{\rm sSFR}>) = a_1 * \log_{10}(1 + \delta_3) + a_0$
  (cf.\ Fig.\ \ref{ssfr_environ_fig}), for all galaxy samples used. For
  details regarding the various galaxy samples, refer to \S \ref{sec_data2}
  and Table \ref{sample_descript_tab}.}
%\label{ssfr_fits_tab}
\end{deluxetable}

\subsection{The SFR--density relation at $z \sim 0.1$ and $z \sim 1$}
\label{sec_results_sfr}

\begin{figure*}[tb]
\centering
\plotone{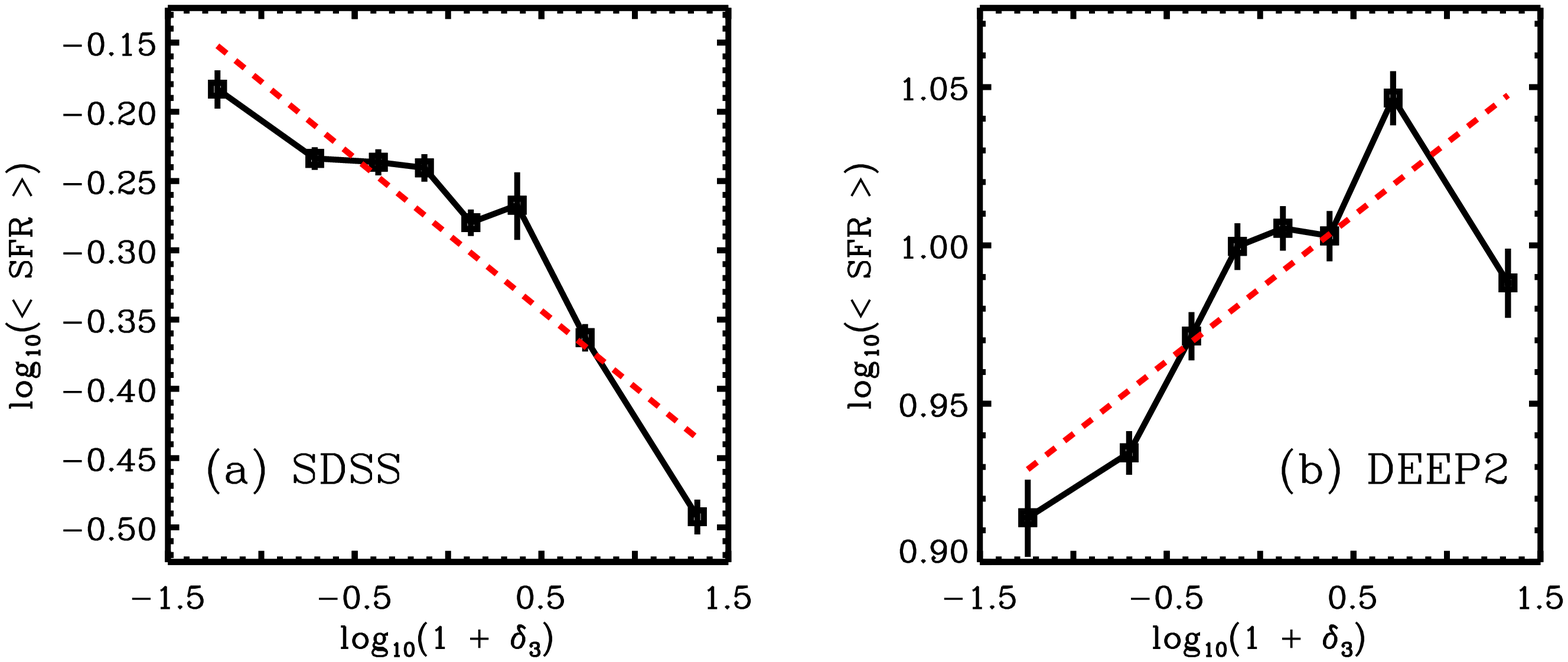}
\caption{The dependence of mean SFR on environment at $z \sim 0.1$
  (\emph{left}) and at $z \sim 1$ (\emph{right}). We plot the logarithm of
  the mean SFR and of the error in the mean SFR in discrete bins of galaxy
  overdensity within the SDSS--A and DEEP2--A samples. The dashed red line
  in each plot shows a linear--regression fit to the data points, with
  coefficients of the fits given in Table \ref{sfr_fits_tab}. Note that the
  star--formation rate (SFR) is given in units of $h^{-2} {\rm M}_{\sun} /
  {\rm yr}$. }
\label{sfr_environ_fig}
\end{figure*}

As shown in Figure \ref{sfr_environ_fig}a, we find that for SDSS--A, the
mean SFR of galaxies in the sample decreases in regions of higher
overdensity, mimicking the sSFR--density relation observed both at $z \sim
0.1$ and at $z \sim 1$. In stark contrast, the mean SFR {\it increases}
with local galaxy density at $z \sim 1$, an inversion of the local relation
(cf.\ Fig.\ \ref{sfr_environ_fig}b). For each of the SDSS and DEEP2
samples, we find similar results to those shown in Figure
\ref{sfr_environ_fig}. Table \ref{sfr_fits_tab} provides the coefficients
from linear--regression fits to the dependence of mean SFR on galaxy
overdensity in each subsample described in \S \ref{sec_data2}. The fits to
the SDSS--A and DEEP2--A samples are illustrated in Fig.\
\ref{sfr_environ_fig} as dashed red lines. While the DEEP2--C sample yields
no detectable correlation between mean SFR and environment, within the
DEEP2--C sample we would have detected the same trend as seen in the
SDSS--C sample at a $\sim \! 10\sigma$ level, given measurement errors.

\begin{deluxetable}{l l l l l}
\tablewidth{0pt}
\tablecolumns{5}
\tablecaption{\label{sfr_fits_tab} Fits to SFR--density Relation}
\tablehead{ Sample & $a_0$ & $a_1$ & $\sigma_{a_0}$ & $\sigma_{a_1}$}
\startdata
DEEP2--A & 0.99 & 0.046 & 0.003 & 0.005 \\
DEEP2--B & 1.07 & 0.034 & 0.003 & 0.005 \\
DEEP2--C & 1.23 & 0.004 & 0.005 & 0.008 \\
SDSS--A & -0.29 & -0.108 & 0.004 & 0.005 \\
SDSS--B & -0.16 & -0.117 & 0.005 & 0.007 \\
SDSS--C & -0.06 & -0.082 & 0.005 & 0.008 \\
\enddata
\tablecomments{We list the coefficients and 1$\sigma$ uncertainties for the
  parameters of least--squares linear--regression fits to the SFR--density
  relation given by $\log_{10}(<{\rm SFR}>) = a_1 * \log_{10}(1 + \delta_3)
  + a_0$ (cf.\ Fig.\ \ref{sfr_environ_fig}), for all galaxy samples
  used. For details regarding the various galaxy samples, refer to \S
  \ref{sec_data2} and Table \ref{sample_descript_tab}.}
%\label{sfr_fits_tab}
\end{deluxetable}

\begin{figure*}[tb]
\centering
\plotone{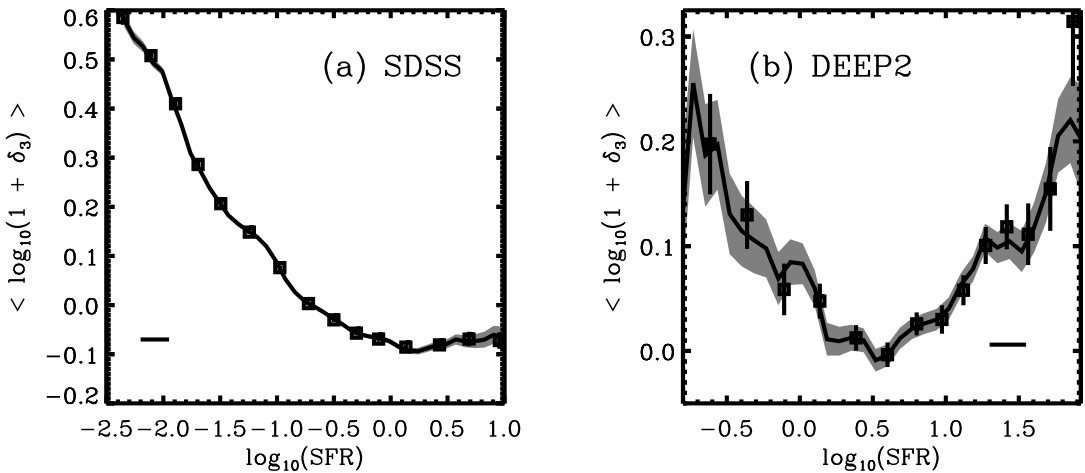}
\caption{The dependence of the mean environment on SFR in the SDSS--A
  (\emph{left}) and DEEP2--A (\emph{right}) samples. We plot the mean and
  the error in the mean of the logarithm of the local galaxy overdensity in
  discrete bins of SFR (square points). The solid black lines show the mean
  dependence of environment on SFR, where the means were computed using
  sliding boxes with widths given by the black dashes in the plot. The
  accompanying grey regions correspond to the sliding 1$\sigma$
  uncertainties in the means. Note that the mean star--formation rate is
  given in units of $h^{-2} {\rm M}_{\sun} / {\rm yr}$.}
\label{environ_sfr_fig}
\end{figure*}

Examining the dependence of mean environment on SFR, we find additional
evidence that the relationship between star--formation activity and local
environment at $z \sim 1$ was dissimilar from that observed at $z \sim
0.1$. In the local Universe, the mean galaxy overdensity smoothly decreases
for galaxy populations with higher star--formation rates, as shown in
Figure \ref{environ_sfr_fig}a. At higher redshift, however, the dependence
of mean overdensity on SFR is considerably more complicated (cf.\ Fig.\
\ref{environ_sfr_fig}b), and is not a simple remapping of Fig.\
\ref{sfr_environ_fig}b. While the mean SFR for galaxies at $z \sim 1$
monotonically increases with increasing overdensity, the dependence of mean
environment on SFR at $z \sim 1$ is a more complex, non--monotonic
relation. This striking difference in the relationship between galaxy
properties and environment at $z \sim 1$ and at $z \sim 0.1$ requires
explication.

\section{Interpreting the Results}
\label{sec_interpret}

To have any hope of accurately characterizing the role of environment in
the cosmic star--formation history, we must understand the differences
between our results at $z \sim 1$ and the corresponding relations at $z
\sim 0.1$. We begin by exploring the relationship between sSFR and
environment, which shows qualitative agreement at low and intermediate
redshifts.

\subsection{Understanding the sSFR--density relation at $z < 1$}

The sSFR--density relations at $z \sim 0.1$ and at $z \sim 1$ can be
explained by the same fundamental physical phenomena that drive the
color--density relation at $z < 1$. As first shown by \citet{cooper06}, the
general form of the color--density relation, as measured locally
\citep{blanton05}, was already established when the Universe was half its
present age, with red galaxies, on average, favoring regions of higher
density relative to their blue counterparts. This strong effect, which is
well studied at low and intermediate redshift \citep[e.g.,][]{hogg03,
  balogh04b, nuijten05, baldry06, weinmann06}, likely results from the
quenching of star formation occurring more efficiently in regions of higher
galaxy density. Many physical processes, from ram--pressure stripping to
galaxy mergers, naturally produce such a connection between the
star--formation history of a galaxy and its local environment \citep[for a
more complete discussion of likely mechanisms, see][]{cooper06}.

\begin{figure*}[tb]
\centering
\plotone{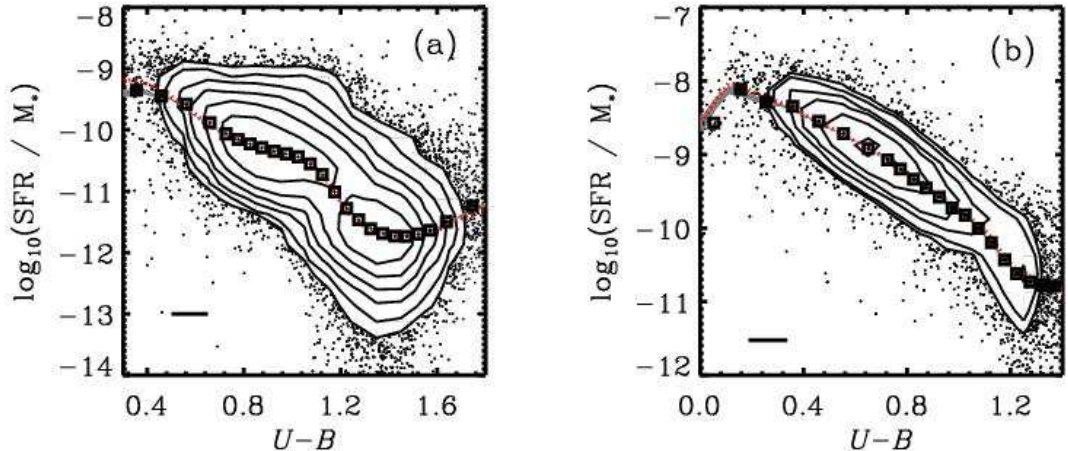}
\caption{The relationship between $\log_{10}({\rm sSFR})$ and rest--frame
  color in the SDSS--A (\emph{left}) and DEEP2--A (\emph{right})
  samples. We plot the mean \emph{square points} and the error in the mean
  \emph{grey region} of $\log_{10}({\rm sSFR})$ in bins of rest--frame
  $U-B$ color. The \emph{red dotted lines} illustrate the median relation
  in the same sliding bins. }
\label{color_ssfr_fig}
\end{figure*}

To illustrate the connection between the sSFR--density and color--density
relations, it is essential to understand the relationship between sSFR and
rest--frame color at $z < 1$. The specific star--formation rate measures
the marginal rate of ongoing star--formation activity in a galaxy. On the
other hand, rest--frame $U-B$ color is a tracer of a galaxy's
star--formation history on roughly Gyr timescales. Although galaxy color
and sSFR measure star--formation activity on different timescales, Figure
\ref{color_ssfr_fig} shows that there is a close relationship between the
two galaxy properties at $z \sim 0.1$ and particularly at $z \sim 1$ (where
$U-B$ colors are better determined). All of the highest--sSFR galaxies are
blue, while redder galaxies have longer star--formation timescales. Because
of the differences in timescales probed by [O {\small II}] emission
($\lesssim 10^7$ years) and the color of stellar populations, the tightness
of this relationship at $z\sim 1$ suggests that the DEEP2 sample is
dominated by galaxies with a smoothly evolving star--formation rate, rather
than undergoing a series of brief, violent star--formation episodes
\citep[see also][]{noeske07a}.

\begin{figure*}[tb]
\centering
\plotone{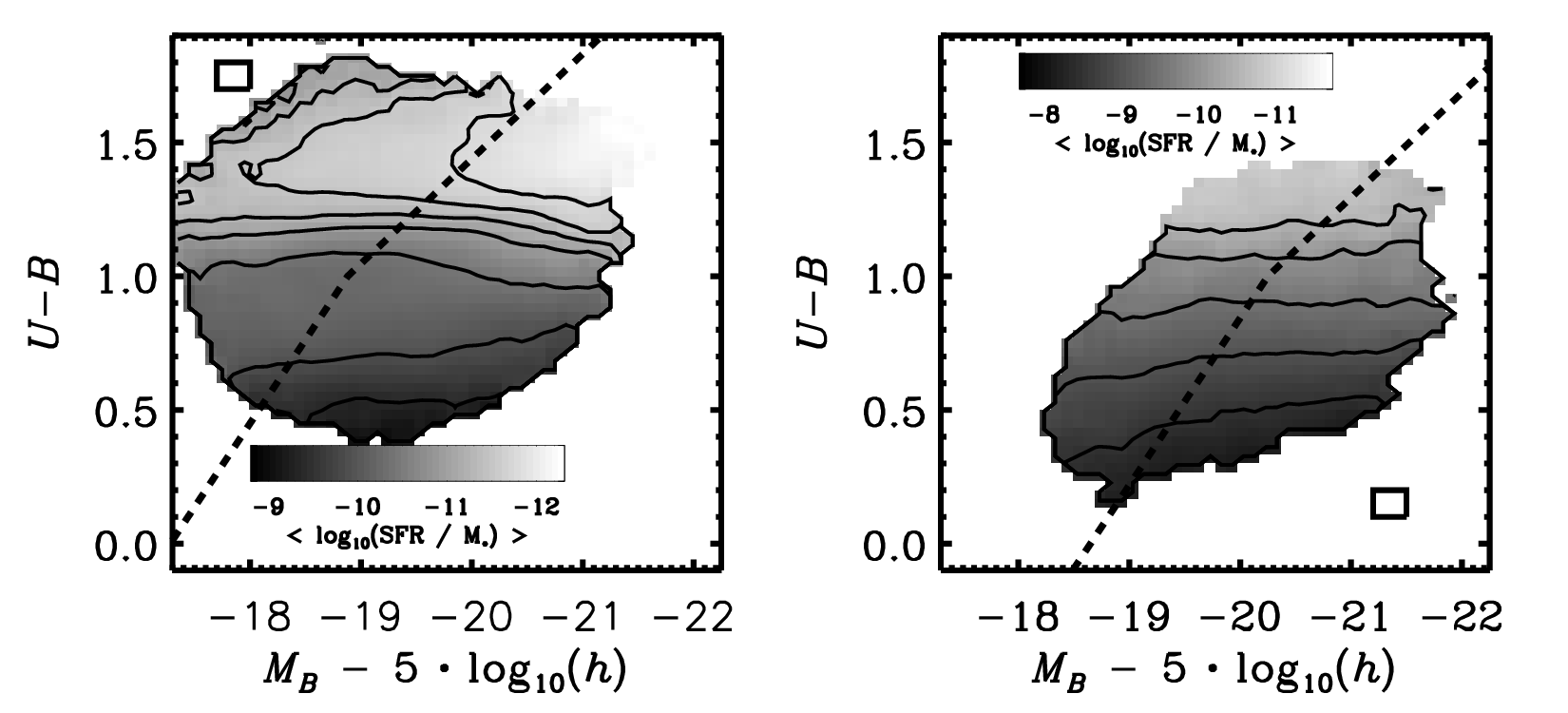}
\caption{The mean sSFR as a function of galaxy color, $U-B$, and absolute
  magnitude, $M_B$, for galaxies in the SDSS--A (\emph{left}) and DEEP2--A
  (\emph{right}) samples. The means are computed in a sliding box,
  illustrated in the corner of each plot, with width $\Delta M_B = 0.3$ and
  height $\Delta (U-B) = 0.1$. Darker areas in the image correspond to
  regions of higher average sSFR in color--magnitude space with the scale
  given by the corresponding inset color bar. The contours in the
  respective plots correspond to levels of $<\log_{10}({\rm sSFR})> =
  -11.75, -11.5, -11.25, -11, -10.5, -10, -9.5$ (\emph{left}) and
  $<\log_{10}({\rm sSFR})> = -10.5, -10, -9.5, -9, -8.5$ (\emph{right}). At
  regions where the sliding box includes less than 20 galaxies, the mean
  sSFR is not displayed. The dashed lines in each plot show the
  color--dependent, absolute--magnitude selection cut (cf.\ Equation
  \ref{magcut_eqn}) used in defining sample DEEP2--B.}
\label{mean_ssfr_cmd}
\end{figure*}

The connection between sSFR and color is even more striking in Figure
\ref{mean_ssfr_cmd}, where we present the mean sSFR as a function of $U-B$
color and absolute magnitude, $M_B$. Both locally and at intermediate
redshift, the mean galaxy sSFR is nearly independent of luminosity at fixed
color. Along the red sequence at $z \sim 0.1$, there is some dependence on
absolute magnitude, such that sSFR is lower in brighter red galaxies. This
magnitude dependence is partially responsible for the greater scatter in
the correlation between sSFR and $U-B$ color in Fig.\ \ref{color_ssfr_fig}a
(relative to Fig.\ \ref{color_ssfr_fig}b); the much larger uncertainties in
SDSS $U-B$ colors, compared to DEEP2, is also important. Galaxies in the
most massive clusters in the SDSS, which are too rare to be probed
significantly by DEEP2 \citep{gerke05}, may also influence the gradient in
mean sSFR along the red sequence at $z \sim 0.1$. A number of physical
processes which are expected only to be significant in such extreme
environments can strip gas from cluster members, thereby cutting off star
formation. Furthermore, clusters are often home to the most massive
red--sequence galaxies, which bias the luminous red galaxy population to
low sSFR. Another source of scatter between specific star--formation rates
derived from [O {\small II}] emission and rest--frame color along the red
sequence is the higher rate of AGN/LINER activity among the red--sequence
population. This point is discussed in more detail in \S \ref{sec_syst2}.

The steepening of the relationship between mean environment and sSFR (cf.\
Fig.\ \ref{environ_ssfr_fig}) at very low specific star--formation rates
can also be understood in terms of the color--density relation. At
$\log_{10}({\rm sSFR}) \lesssim -10.5\ {\rm yr}^{-1}$ in the SDSS and
$\log_{10}({\rm sSFR}) \lesssim -9\ {\rm yr}^{-1}$ in DEEP2, the
sSFR--density relation increases in strength such that the mean overdensity
rises dramatically for galaxies with longer star--formation timescales. At
these very low specific star--formation rates, the SDSS and DEEP2 samples
transition from being dominated by blue galaxies to members of the red
sequence (cf.\ Fig.\ \ref{color_ssfr_fig}). As shown in Figure 5 of
\citet{cooper06} and Figure 2 of \citet{blanton05}, the average overdensity
exhibits a similarly sharp rise at the transition from the blue cloud to
the red sequence. This relatively strong change from a blue galaxy
population with short star--formation timescales typically residing in
environments near the mean cosmic density, to a red population with low
specific star--formation rates commonly located in higher--density regions,
indicates that local environment plays a central role in the truncation of
star formation in galaxies at $z < 1$.

Considering previous studies of galaxy properties at $0 < z < 1$, the close
connection between environment, color, and sSFR is not particularly
surprising. From a study of [O {\small II}] equivalent width, which roughly
traces sSFR, in clusters and field samples at slightly lower redshifts
$(0.18 < z < 0.55)$, \citet{balogh98} arrived at a similar result, finding
that the typical [O {\small II}] equivalent width for a galaxy decreases as
a function of clustercentric radius. Furthermore, using data from the DEEP2
survey to study galaxy environments at $z \sim 1$, \citet{cooper06} found
that [O {\small II}] equivalent width (which is closely correlated with
sSFR) is anticorrelated with local galaxy density, on average.

\subsection{Understanding the SFR--density relation at $z < 1$}

\begin{figure*}[tb]
\centering
\plotone{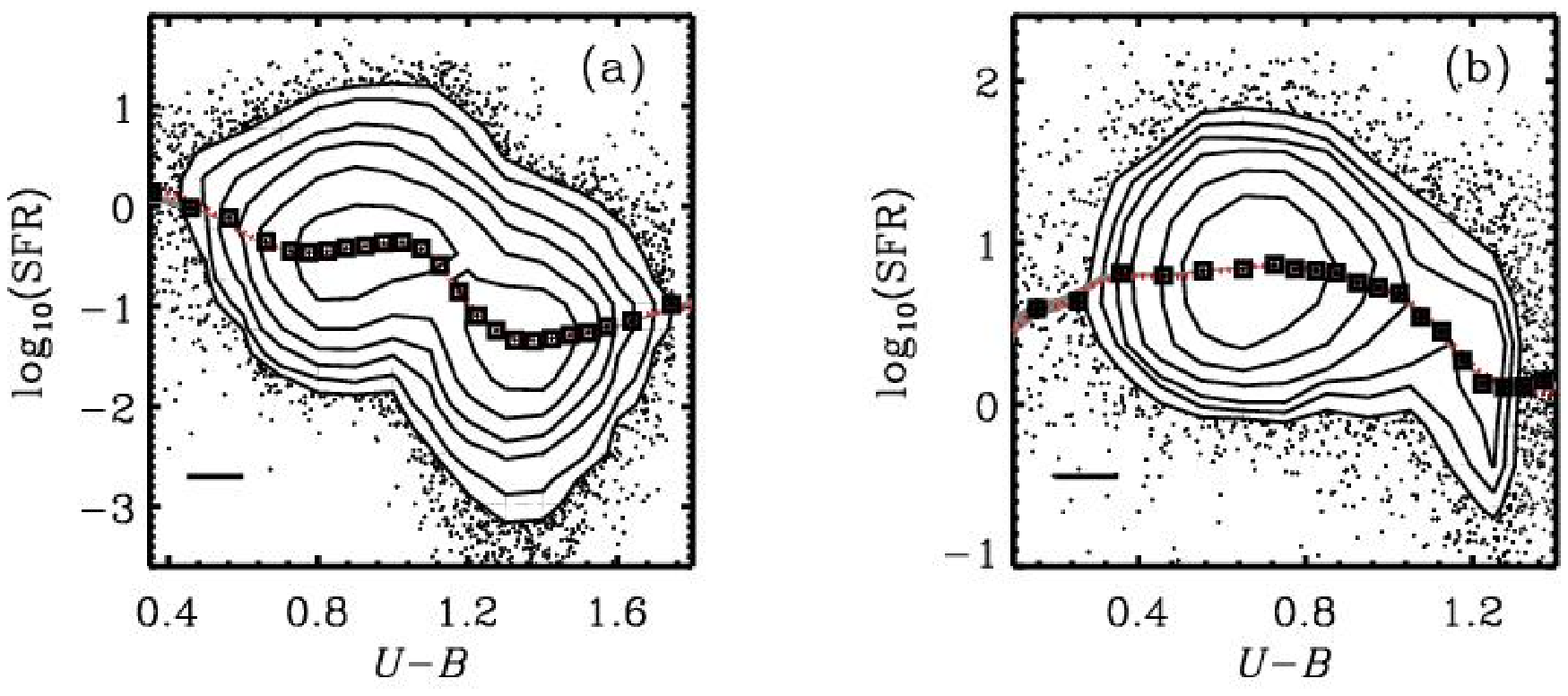}
\caption{The relationship between $\log_{10}({\rm SFR})$ and rest--frame
  color in the SDSS--A (\emph{left}) and DEEP2--A (\emph{right})
  samples. We plot the mean \emph{square points} and the error in the mean
  \emph{grey region} of $\log_{10}({\rm SFR})$ in bins of rest--frame
  $U-B$ color. The \emph{red dotted lines} illustrate the median relation
  in the same sliding bins.}
\label{color_sfr_fig}
\end{figure*}

As discussed above, the specific star--formation rates in nearby galaxies
are closely tied to their rest--frame $U-B$ colors. For {\it total} SFR,
however, the connection to rest--frame color is not nearly as simple.  In
Figure \ref{color_sfr_fig}, we show the relationship between total SFR and
galaxy color at $z \sim 0.1$ and at $z \sim 1$. At constant color, the
range of star--formation rates can exceed two orders of magnitude.

\begin{figure*}[tb]
\centering
\plotone{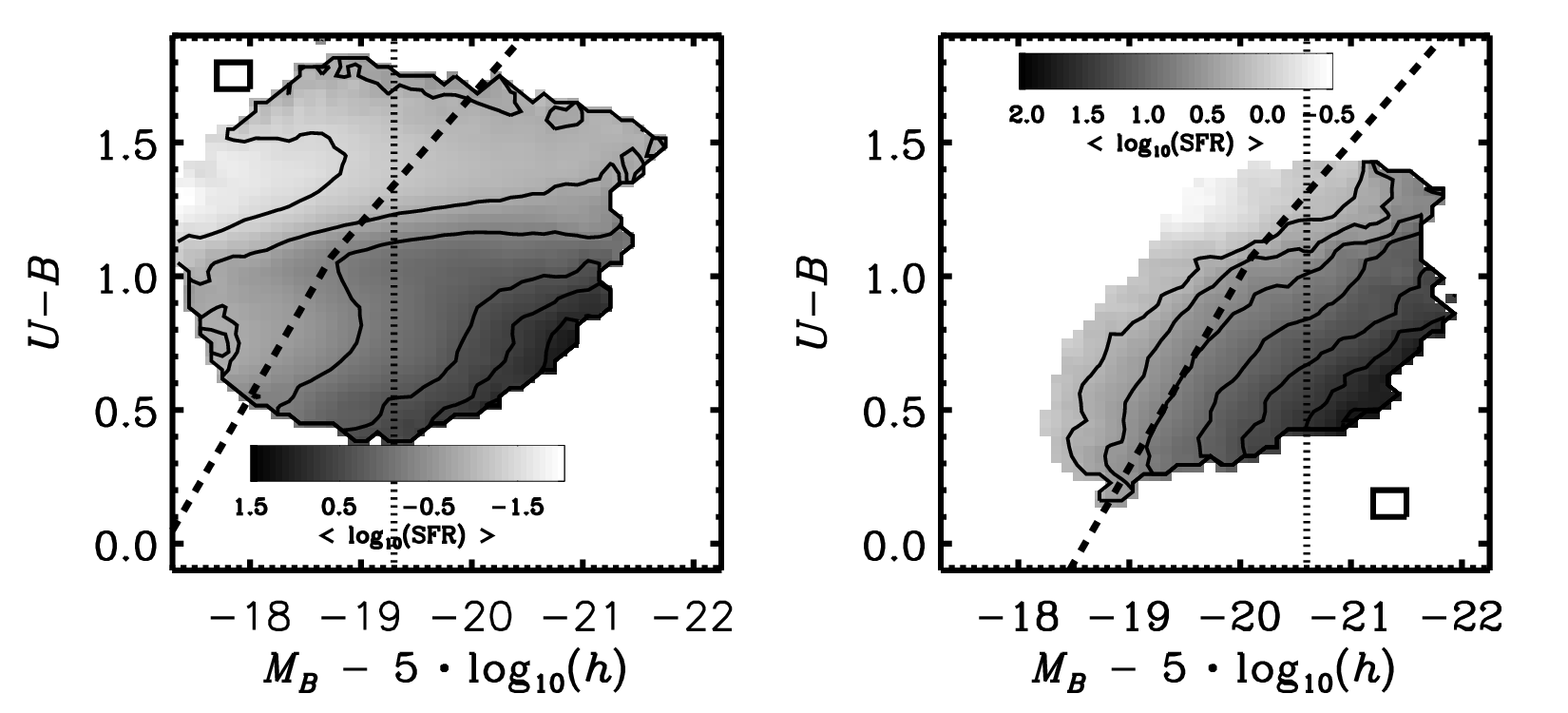}
\caption{The mean SFR as a function of galaxy color, $U-B$, and absolute
  magnitude, $M_B$, for galaxies in the SDSS--A (\emph{left}) and DEEP2--A
  (\emph{right}) samples. The means are computed in a sliding box,
  illustrated in the corner of each plot, with width $\Delta M_B = 0.3$ and
  height $\Delta (U-B) = 0.1$. Darker areas in the image correspond to
  regions of higher average star--formation activity in color--magnitude
  space with the scale given by the corresponding inset color bar. The
  contours in the respective plots correspond to levels of $<\log_{10}({\rm
    SFR})> = -1.5, -1, -0.5, 0.0, 0.25, 0.5$ (\emph{left}), and
  $<\log_{10}({\rm SFR})> = 0.25, 0.5, 0.75, 1, 1.25, 1.5, 1.75$
  (\emph{right}). At regions where the sliding box includes less than 20
  galaxies, the mean SFR is not displayed. The dashed lines in each plot
  show the color--dependent, absolute--magnitude selection cut (cf.\
  Equation \ref{magcut_eqn}) used in defining sample DEEP2--B.}
\label{mean_sfr_cmd_fig}
\end{figure*}

Though there is an overall trend between SFR and $U-B$ color, the
correlation is much weaker than that seen between sSFR and $U-B$ color
(cf.\ Fig.\ \ref{color_ssfr_fig}). The factor of stellar mass which
distinguishes sSFR from SFR is the logical culprit. When we examine the
mean SFR as a function of both $U-B$ color and absolute magnitude, $M_B$,
we find that, unlike the mean sSFR, the mean SFR is far from independent of
$M_B$; in Figure \ref{mean_sfr_cmd_fig}, the isocontours of mean SFR run
diagonally.\footnote{The difference in the slope of the isocontours of mean
  SFR in the SDSS and DEEP2 samples is discussed in more detail in \S
  \ref{sec_syst3}.} This is little surprise, since sSFR is closely related
to galaxy color, as shown in Fig.\ \ref{color_ssfr_fig}, and $({\rm
  M}_{*}/L)$ is to first order simply a function of color, as described in
\S \ref{sec_data5}; so total SFR is, roughly, a product of two functions of
color alone (sSFR and $({\rm M}_{*}/L_B)$ with the luminosity defined by
$M_B$. Thus, evolution in the relationship between global SFR and
environment is closely tied to evolution in the relationships between $U-B$
and environment as well as $M_B$ and environment, not just one or the
other.

\begin{figure}[h]
\centering
\plotone{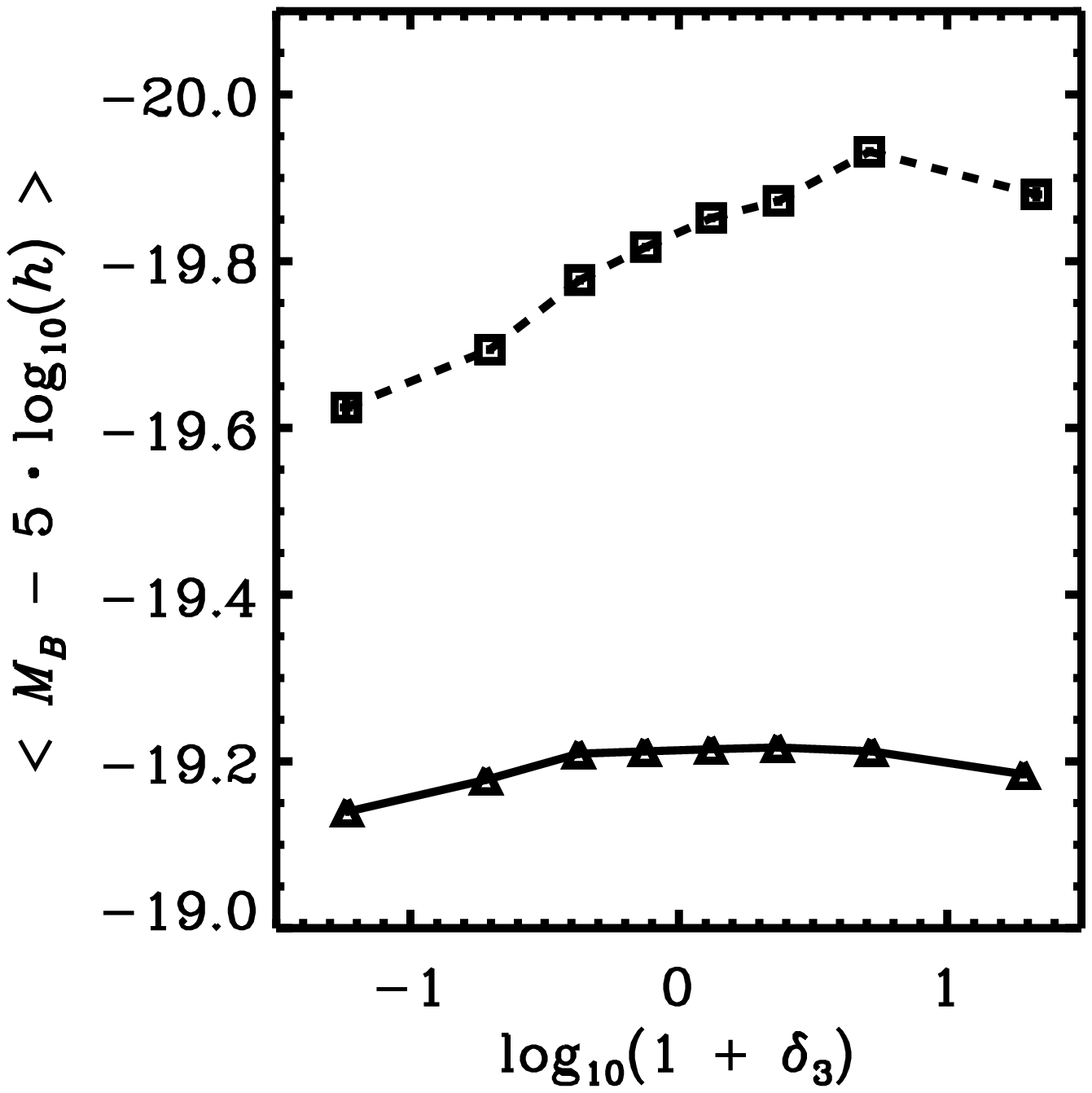}
\caption{The mean absolute $B$--band magnitude, $M_B$ for blue
  galaxies in the SDSS--A (\emph{solid line and triangles}) and DEEP2--A
  (\emph{dashed line and squares}) samples, in bins of local galaxy
  overdensity, $1 + \delta_3$. The blue galaxy populations within the SDSS
  and DEEP2 samples are defined as described in \S \ref{sec_data1}. While
  the mean $M_B$ shows little dependence on environment for nearby blue
  galaxies, there is a much stronger dependence at $z \sim 1$.}
\label{mag_environ_fig}
\end{figure}

Using data from the DEEP2 survey, \citet{cooper06} explored the dependence
of mean environment on the two dimensions of galaxy color and absolute
magnitude at $z \sim 1$. One of the chief results from this work was the
discovery that for galaxies in the blue cloud the mean galaxy environment
depends on $M_B$. In contrast, previous studies at $z \sim 0$ found that
mean galaxy overdensity is essentially independent of absolute magnitude
for the blue galaxy population \citep[e.g.,][]{hogg04, blanton05}. In
Figure \ref{mag_environ_fig}, we show the mean absolute $B$--band magnitude
as a function of local overdensity for blue galaxies in the SDSS and DEEP2
survey; while the mean $M_B$ among nearby galaxies varies only weakly as a
function of $1 + \delta_3$, the same trend at $z \sim 1$ shows a
significant gradient, such that the average absolute magnitude of galaxies
in regions of higher galaxy density is brighter. From linear--regression
fits to the data points in Fig.\ \ref{mag_environ_fig}, we find that the
slope of the relationship between mean absolute magnitude ($M_{B}$) and
environment for blue galaxies is a factor of $> \! 5$ greater at $z \sim 1$
than that found locally.

This difference in the relationship between environment and absolute
magnitude (and therefore stellar mass --- cf.\ Fig.\ \ref{mean_mass_cmd})
within the blue galaxy populations at $z \sim 0.1$ and $z \sim 1$ is
reflected in the observed inversion in the mean SFR--environment
relationship. In combination with the color--density relation,
it can explain the trends found in both Fig.\ \ref{sfr_environ_fig} and
Fig.\ \ref{environ_sfr_fig}. When plotting the mean SFR as a function of
environment at $z \sim 1$, each bin in overdensity is dominated by galaxies
residing in the blue cloud, given DEEP2's bias towards the blue galaxy
population. Thus, in Fig.\ \ref{sfr_environ_fig}b, we simply see an
increase in mean SFR with overdensity, tracking the trend between mean
$M_{B}$ and environment for blue galaxies (cf.\ Fig.\
\ref{mag_environ_fig}). At the highest galaxy densities, the contribution
of red galaxies to the DEEP2 sample peaks and contributes to the flattening
of this relationship.

In Fig.\ \ref{environ_sfr_fig}, on the other hand, we see a U--shaped trend
when plotting mean environment as a function of SFR at $z \sim 1$. At the
lowest SFRs, the sample is dominated by the red galaxy population (cf.\
Fig.\ \ref{mean_sfr_cmd_fig}), which are quenched entirely. Thus, we
measure a high average overdensity for galaxies with low levels of
star--formation activity, driven by the color--density relation; red
galaxies typically favor regions of higher galaxy density, at both $z\sim
0.1$ and $z\sim 1$. At somewhat higher SFRs, the sample becomes a mix of
galaxies on the red sequence and fainter galaxies on the blue cloud, so the
mean environment drops, reflecting the color--density relation. At
$\log_{10}({\rm SFR}) \gtrsim 0.8\ {\rm M}_*/{\rm yr}$, however, the mean
overdensity begins to rise as the sample becomes entirely composed of
increasingly brighter blue, star--forming galaxies. At these high SFRs, we
therefore find the same relation as seen in Fig.\ \ref{sfr_environ_fig}b,
where the mean overdensity increases with SFR.

%%%%%%%%%%%%%%%%%%%%%%%%%%%%
%%% 5: Selection Effects %%%
%%%%%%%%%%%%%%%%%%%%%%%%%%%%
\section{Potential Systematic Effects}
\label{sec_syst}

\subsection{Isolating the Blue Cloud}
\label{sec_syst1}

As briefly discussed in \S \ref{sec_data}, the galaxy population at $z
\lesssim 1$ is bimodal in nature, with galaxies both locally and out to
intermediate redshifts divisible into two distinct types: red, early--type
galaxies lacking much star formation and blue, late--type galaxies with
active star formation. Separating these two galaxy populations as described
in \S \ref{sec_data1}, we can study the blue (``star--forming'')
population in isolation while eliminating any relative bias towards one or
the other type of galaxy in the SDSS and DEEP2 samples.  The SDSS--C
sample, which has a red fraction matched to that of the DEEP2--C sample,
also provides a test for such selection effects. Additionally, this can
facilitate comparison to other environment studies which only include
star--forming galaxies \citep[e.g.,][]{elbaz07}.

We find that the trends of sSFR and SFR with environment discussed in \S
\ref{sec_results} at $z \sim 0$ and $z \sim 1$ persist if we restrict
samples to only the blue, star-forming population. Table \ref{fits2_tab}
lists the coefficients from linear--regression fits to the dependence of
mean SFR (and sSFR) on overdensity in the SDSS--A and DEEP2--A galaxy
samples, isolating simply the blue--cloud population. For both the SDSS and
DEEP2 samples, the dependence of mean sSFR on local overdensity is weaker
when excluding the red sequence. As discussed in \S \ref{sec_interpret},
the sSFR--density relation is effectively tracing the color--density
relation, which is dominated by the tendency of red galaxies -- which have
been removed here -- to be found in dense regions.

When measuring the dependence of mean SFR on local galaxy density, we find
that the relation is weaker in the SDSS blue--cloud population than in the
full SDSS--A sample. However, we still detect a significant trend, with the
mean SFR of the star-forming population being lower in regions of higher
overdensity. Again, this relationship is closely connected to the
color--density relation, and so we would expect it to be weakened by the
exclusion of the red--sequence population. In contrast, we find a stronger
dependence of mean SFR on environment in DEEP2 when isolating the blue
cloud than not. As detailed in \S \ref{sec_interpret}, the relationship
between SFR and galaxy density at $z \sim 1$ is a combination of two
competing trends: the color--density relation and the dependence of mean
luminosity on environment along the blue cloud. The inversion in the
SFR--density relation at $z \sim 1$ relative to $z \sim 0$ is driven by the
latter trend. Isolating the blue cloud effectively minimizes the role of
the color--density relation, making the SFR--density relation monotonic.

\begin{deluxetable*}{l c c c c c c c c l}
\tablewidth{0pt}
\tablecolumns{10}
\tablecaption{\label{fits2_tab} Fits to (s)SFR--density Relations}
\tablehead{Sample & \multicolumn{4}{c}{sSFR--density fits} &
  \multicolumn{4}{c}{SFR--density fits} & Subset Selected \\
  & $a_0$ & $a_1$ & $\sigma_{a_0}$ & $\sigma_{a_1}$ & 
$a_0$ & $a_1$ & $\sigma_{a_0}$ & $\sigma_{a_1}$ & }
\startdata
DEEP2--A & -8.76 & -0.023 & 0.013 & 0.021 & 
1.03 & 0.068 & 0.003 & 0.005 & only blue galaxies \\
DEEP2--A & -8.89 & -0.047 & 0.035 & 0.053 & 
0.79 & 0.046 & 0.006 & 0.010 & with AGN removed \\

SDSS--A & -9.99 & -0.038 & 0.023 & 0.034 &
-0.07 & -0.007 & 0.004 & 0.006 & only blue galaxies \\ 
SDSS--A & -10.13 & -0.120 & 0.026 & 0.040 & 
-0.79 & -0.107 & 0.004 & 0.006 & with AGN removed \\

\enddata
\tablecomments{We list the coefficients and 1$\sigma$ uncertainties for the
  parameters of the linear--regression fits to the sSFR--density and
  SFR--density relations given by $\log_{10}(<{\rm (s)SFR}>) = a_1 *
  \log_{10}(1 + \delta_3) + a_0$ (cf.\ Fig.\ \ref{ssfr_environ_fig} and
  Fig.\ \ref{sfr_environ_fig}), for subsamples of the SDSS--A and DEEP2--A
  galaxy samples selected (i) to isolate the blue cloud and (ii) to remove
  AGN. For details of these selections, refer to \S
  \ref{sec_syst1} and \S \ref{sec_syst2}.}
%\label{fits2_tab}
\end{deluxetable*}

\subsection{The Impact of AGN Contamination}
\label{sec_syst2}

Another sensible reason for focusing solely on the blue--cloud population
is to minimize the impact of contamination from active galactic nuclei (or
AGN). As shown and discussed in detail by \citet{yan06}, [O {\small II}]
$\lambda3727$\AA\ emission is not a direct indicator of star formation in
the local Universe. AGN, especially LINERs \citep{heckman80}, can
contribute significantly to the integrated [O {\small II}] emission from
nearby galaxies. This additional source of emission causes the measured SFR
derived from [O {\small II}] line luminosities to be an overestimate of the
true total SFR for some portion of the galaxy population. The impact of
such AGN activity on the galaxy sample is greatest along the red sequence;
where roughly 35--45\% of nearby red galaxies exhibit signatures of AGN
activity \citep{yan06}.

\citet{yan06} show that galaxies with significant AGN activity can be
identified based on emission--line diagnostics (e.g., [O {\small
  II}]/H$\alpha$, [N {\small II}]/H$\alpha$, and H$\beta$/[O {\small III}]
line ratios), and therefore removed from the galaxy sample \citep[see
also][]{kewley06}. In the SDSS, we identify galaxies with likely AGN
contamination based on several line--ratio criteria:
\begin{equation}
\rm{[N\ {\small II}]} / \rm{H}\alpha > 0.6 
\label{yan_cut1}
\end{equation}
\begin{equation}
W_{\rm [O\ II]} > 5 \cdot W_{{\rm H}\alpha} - 7
\label{yan_cut2}
\end{equation}
\begin{equation}
  \log_{10}{(\rm{[O\ III]} / \rm{H}\beta)} > 0.61\ /\ 
  [\log_{10}{(\rm{[N\ II]}/\rm{H}\alpha)} - 0.05 ] + 1.3
\label{yan_cut3}
\end{equation}
where $W_{x}$ and $x$ denote equivalent widths and line luminosities for a
given line, respectively. Equation \ref{yan_cut1} is designed to select
most LINERs and a majority of Seyferts, while equation \ref{yan_cut2} is
effective at identifying nearly all LINERS. Finally, equation
\ref{yan_cut3} is based on AGN--selection in the SDSS by
\citet{kauffmann03}, which is less sensitive to aperture effects related to
the SDSS fibers and therefore more inclusive in selecting AGN than the
criterion proposed by \citet{kewley01}. Any source meeting any of these
criteria is flagged as a likely AGN and subsequently removed from the
galaxy sample. The union of these three selection criteria gives an
extremely inclusive sample of AGN, with many AGN--star--forming composites
also being removed from the main galaxy sample as AGN.

At $z \sim 0.1$, we conclude that contamination from AGN does not introduce
any systematic bias that strongly impacts the observed relationships
between star formation and environment. Table \ref{fits2_tab} lists the
coefficients from linear--regression fits to the dependence of mean SFR
(and sSFR) on local galaxy overdensity for the SDSS--A sample with the AGN
population removed (but non--AGN red galaxies still retained). For this
restricted galaxy sample, we find that the results detailed in \S
\ref{sec_results} for the full SDSS galaxy sample persist, with mean SFR
and mean sSFR decreasing in regions of higher galaxy density. The strength
of the relationship between average sSFR and environment is weaker when the
AGN population is removed, largely due to the significantly smaller number
of red galaxies in the SDSS sample after removing AGN contamination (again,
reflecting the fact that the concentration of red galaxies in overdense
regions dominates the color--density relation).

At $z \sim 1$, H$\alpha$ is redshifted out of the optical window, and thus
out of the spectral range probed by DEEP2. As a result, the [O {\small II}]
emission from AGN is more difficult to disentangle from star--formation
activity in our higher--redshift galaxy sample. While at $0.74 \lesssim z
\lesssim 0.83$ we are able to use [O {\small II}], H$\beta$, and [O {\small
  III}] line ratios to distinguish AGN--like emission \citep{yan06, yan07},
restricting to galaxies in such a small redshift window severely constricts
the sample size at high redshift ($< \! 3500$ galaxies). For this small
sample, we find no significant difference in the relationships between
sSFR, SFR, and environment with those found in the full DEEP2--A sample
(cf.\ Table \ref{fits2_tab}). We note that the exclusion of AGN in DEEP2 is
only effective at removing Seyfert and LINER--like AGN emission; transition
objects (TOs) tend to have similar [O {\small III}]/H$\beta$ ratios to
those of the star--forming galaxies. To confidently exclude transition
objects, [N {\small II}]/H$\alpha$ line diagnostics are needed
\citep{yan07}. However, transition objects comprise only a fraction of the
AGN--like population (e.g., $\sim \! 10\%$ of red galaxies in the SDSS
are TOs).

For these reasons, we are not able to definitively gauge the impact of AGN
activity on our results at $z \sim 1$, other than to show it should not
dominate the observed trends. As previously discussed, the contribution of
AGN--related emission to [O {\small II}] line luminosities produces
overestimated SFRs. However, as shown by \citet{nandra07} from X--ray
observations of the Extended Groth Strip, most AGN at $z \sim 1$ are either
along the red sequence or in between the red sequence and blue cloud,
similar to what has been observed locally \citep{yan06, yan07}. When we
restrict to just galaxies in the blue cloud within DEEP2, the inversion in
the local SFR--density relation still persists, with SFR increasing in
regions of higher density at $z \sim 1$; AGN cannot be driving this
inversion.

In addition, spectroscopically--identified AGN comprise only a small
portion $(< \! 5\%)$ of the DEEP2 sample at moderate to high
star--formation rates $({\rm SFR} > 10\ {\rm M}_{\sun}/{\rm yr})$. As such,
AGN cannot be responsible for the rise in mean overdensity at high SFRs in
Figure \ref{environ_sfr_fig}. Instead, AGN contamination would cause the
U--shaped trend in Fig.\ \ref{environ_sfr_fig} to be flattened, as red
galaxies with no star--formation activity (which preferentially reside in
regions of high galaxy density) would be placed into the range $0.25 <
\log_{10}({\rm SFR}) < 1$ due to misaccounting of their AGN activity. Thus,
we conclude that AGN contamination in the DEEP2 data set is not a viable
explanation for the observed inversion in the SFR--density relation at $z
\sim 1$.

Still, the impact of AGN contamination is clearly a concern when studying
star formation at higher redshift, especially given the higher number
density of AGN (both low-- and high--luminosity) at $z \sim 1$ relative to
what is found at $z \sim 0$ \citep[e.g.,][]{cowie03, hasinger03,
  barger05}. A more detailed treatment of the role of AGN in DEEP2 will be
included in future works using the multiwavelength data in the EGS
\citep{yan07, woo07, marcillac07, georgakakis07}.

\subsection{The Role of Dust}
\label{sec_syst3}

In contrast to AGN activity, which can lead to over--inflated SFR
estimates, dust will cause some fraction of star formation to be hidden at
optical or UV wavelengths, yielding a measurement of the total SFR that is
an underestimate of the true SFR. To account for this effect, we have
corrected our SFRs according to the mean relations of \citet{moustakas06},
which calibrate SFRs derived from [O {\small II}] line luminosities against
more precise SFR indicators (e.g., extinction--corrected H$\alpha$--derived
SFRs) from multiwavelength data; they find it is possible to produce a
correction which is a function only of galaxy luminosity, $M_B$, that
accounts for both mean metallicity and reddening effects. While
\citet{moustakas06} conclude that there remains significant scatter between
these [O {\small II}]--derived SFR estimates and cleaner measures of star
formation (at a level of roughly $\pm 90\%$), this lack of precision in our
star--formation rates is not the dominant source of error in our
analysis. Measures of galaxy environment are by nature imprecise, with a
level of uncertainty dependent upon the sampling density and redshift
precision and accuracy of the redshift survey. The errors in overdensity
measures are the greatest obstacle to detecting trends between galaxy
properties and environment at $z \sim 1$ and at $z \sim 0$.  By applying
the corrections of \citet{moustakas06}, we should be able to reduce any
overall bias in our analysis due to the use of [O {\small II}] as a SFR
indicator to a modest level.

While the empirical calibrations of \citet{moustakas06} were derived from
data at $z \sim 0$, they have been tested at higher redshifts (to $z
\lesssim 1.4$), with no dependence on redshift observed. Still, the DEEP2
sample employed here probes fainter luminosities than the samples used by
\citet{moustakas06} at $z > 0.7$. To test the sensitivity of our results to
the particular calibration of the [O {\small II}]--derived star--formation
rates at $z \sim 1$, we have redone our analysis with star--formation rate
calibrations having $1/2 \times$ or $2 \times$ the luminosity--dependence
given in Table 2 of \citet{moustakas06}, accounting for a wide range in
evolution of possible dust and metallicity effects. Even when we vary the
luminosity--dependence of the [O {\small II}]/SFR calibration to such a
degree, the form of the trends between both specific and total
star--formation rate and environment at $z \sim 1$ remain similar (e.g.,
the SFR--density relation is still inverted relative to that found at $z
\sim 0$, and the U-shaped dependence of average overdensity on SFR
persists). We note that the {\it amplitude} of this calibration, rather
than the strength of its luminosity--dependence, will not affect the
conclusions of any of our studies at constant redshift, but only
comparisons of the overall strength of star--formation at $z\sim 1$ to
$z\sim 0.1$.

Clearly, if a source is sufficiently highly obscured, then we may detect no
[O {\small II}] emission in the DEEP2 spectra. For such sources, the
empirical calibrations of \citet{moustakas06} will not be
appropriate. However, multiwavelength studies of star formation at $z
\lesssim 3$ have shown that extinction is well--correlated with
star--formation rate \citep{ahopkins01}. As such, correcting for extinction
beyond that expected from \citet{moustakas06} would be expected to increase
the strength of the trends observed in the DEEP2 sample, causing the
average SFR to be yet higher in overdense regions.

One method for detecting heavily obscured sources and minimizing the impact
of dust on SFR measurements is to observe at longer wavelengths (e.g., at
24$\mu$m), where dust obscuration is much weaker. Analysis of the
relationship between star formation and environment using {\it
  Spitzer}/MIPS data will be the subject of future papers
\citep{marcillac07, woo07} and will be a critical check to the results
presented here. However, extremely deep infrared data is needed to include
sources with star--formation rates as low as those measurable in the DEEP2
spectroscopy. For instance, within the EGS, 5421 of the DEEP2 sources are
in the {\it Spitzer}/MIPS region, but only 1645 have 24$\mu$m detections to
a flux limit of $S \sim 83$mJy \citep{weiner07}. Ongoing, extremely deep
{\it Spitzer}/MIPS observations in the EGS and the Extended Chandra Deep
Field South (ECDFS) will provide data sets well--suited to future analyses
of obscured star formation at intermediate redshift.

We note that although they do not appear to influence our environment
results, the effects of dust extinction and AGN contamination may be
responsible for the difference in the curvature of the isocontours of mean
SFR as a function of color and absolute magnitude in the SDSS and DEEP2
samples, as seen in Figure \ref{mean_sfr_cmd_fig}. The higher number
density of AGN at $z \sim 1$ relative to $z \sim 1$, especially on the red
edge of the blue cloud and on the red sequence, inflates the mean SFR at
those portions of the color--magnitude diagrams, thereby stretching the
isocontours of mean SFR to redder colors.  Similarly, the higher fraction
of obscured star formation at intermediate redshift \citep{lefloch05} could
be responsible for elongating the iscontours at $z \sim 1$ with respect to
those at $z \sim 0$.

\section{The Role of Environment in the Cosmic SFH}
\label{sec_disc}

As discussed in \S \ref{sec_intro}, one explanation which has been posited
for the dramatic drop in the cosmic star--formation rate space density at
$z < 1$ is a corresponding steep decline in the galaxy merger rate, which
thereby reduces the amount of merger--induced star formation. Mergers of
dark matter halos are integral to hierarchical structure formation; their
rate may be calculated straightforwardly in the Extended Press--Schechter
formalism \citep{press74, lacey93}. Unfortunately, attempts to constrain
the evolution in the galaxy merger rate through observations at $z < 1$,
have produced mixed results.

Early imaging studies of galaxies at intermediate redshift using the {\it
  Hubble Space Telescope} ({\it HST}) found galaxies with disturbed
morphologies (i.e., mergers) to be much more common at $z \sim 1$ than
locally and attributed the rapid decline in the blue luminosity density to
the evolution of this population \citep[e.g.,][]{abraham96, brinchmann98,
  vdb00}. In agreement with this finding, some estimates of the galaxy
merger rate at $z < 1$ derived from analyses of galaxy morphologies and
from studies of spectroscopic pairs, have found a significant decline in
the frequency of major mergers over the last 7 Gyr \citep{zepf89, patton97,
  lefevre00, conselice03, hammer05, kampczyk07}.

However, a number of other efforts to study galaxies at $z \sim 1$ have
found a much more gradual decline in the pair fraction and merger rate
since $z \sim 1$, downplaying the role of galaxy interactions in the
evolution of the cosmic star--formation history
\citep[e.g.,][]{neuschaefer97, carlberg00, bundy04, lin04, lotz07}. The
abundance of contrasting results is, at least in part, due to the
difficulty in measuring the merger rate, which depends significantly on the
treatment of projection effects and assumptions regarding the timescales
over which merger signatures are visible. Furthermore, despite having large
differences in the best--fit rate, many of the seemingly--contradictory
evolution estimates are still consistent with each other due to small
sample sizes and large random errors. Many systematic problems could affect
one method or another; for instance, \citet{bell06} argues that some of the
variance in results is attributable to accounting errors in converting pair
fractions into merger rates.

Recent studies at intermediate redshift have benefited from significantly
larger sample sizes and deeper {\it HST} imaging over wider fields. While
this has not yielded consensus on the evolution of the merger rate, studies
of galaxy morphologies using data sets in fields such as the EGS and the
ECDFS have found that the contribution from irregular (or peculiar)
galaxies to the blue luminosity density to be subdominant at $z \sim 1$;
for instance, results from \citet{conselice05}, \citet{wolf05}, and
\citet{zamojski07} all indicate that the drop in the global SFR at $z < 1$
(as traced by the $B$--band and UV luminosity densities) is driven by a
decline in the emission from galaxies with regular (i.e., non--disturbed)
morphologies. Thus, regardless of the evolution in the merger rate, galaxy
mergers must not be the dominant cause of the stark \citep[$\sim \!
10$--$20\times$,][]{fukugita03, ahopkins06} decline in the global
star--formation rate since $z \sim 1$.

This conclusion is supported by a variety of other galaxy studies at $z <
1$. \citet{lin07} placed an upper limit of $< \! 40\%$ on the contribution
from galaxy pairs (i.e., precursors to mergers) to $24\mu$m infrared
emission at intermediate redshift. Similarly, \citet{bell05} find that the
relationship between galaxy morphology and star--formation activity at $z
\sim 0.7$ implies that major mergers (and the evolution in the major merger
rate) cannot be the proximate cause of the decline in the cosmic
star--formation rate. In agreement with these observational results,
theoretical models of galaxy mergers predict that such interactions
contribute only a small fraction of the total star--formation rate space
density of the Universe \citep{hopkins06b}.

Since galaxy mergers do not appear to be driving the global decline in the
cosmic star--formation rate since $z < 1$, a new paradigm is required.
Several multiwavelength studies of star formation at intermediate and low
redshifts have concluded that the reduction in the global star--formation
rate space density is largely driven by long--term processes such as
gradual gas depletion, rather than galaxy mergers. For instance,
\citet{bauer05} and \citep{noeske07a} find that star--forming galaxies
populate a tight sequence in SFR and stellar mass at $z < 1.2$, with a
limited range in SFR at a given stellar mass and redshift
\citep{noeske07a}. This constrains the amount of episodic star--formation
activity that may be occurring, and indicates that star formation at $z <
1$ appears to be dominated by a gradual decline in the average SFR among
the star--forming galaxy population. Strong, outlier bursts indicative of
being driven by merger activity are rare, even at $z > 1$
\citep{noeske07b}.

While mergers (and environment) should play some role in the decline of the
global SFR even if there is only a weak evolution in the merger rate, the
growing consensus appears to be that mergers have relatively little effect
on the global decline in star formation from $z \sim 1$ to $z \sim 0$. The
impact of other environment--dependent mechanisms, however, could be
significant. For instance, virial shock--heating of infalling gas onto
massive dark matter halos can produce a smooth decline in star formation
with time: as the flow of cold gas to galaxies in such halos is suspended,
the galaxies suffer from starvation and gradually stop forming stars. Given
the strong correlation between halo mass and local galaxy density
\citep[e.g.,][]{lemson99, gao04, wetzel07, croton07}, this process should
exhibit a close connection with galaxy environment. Similarly, conductive
heating of gas can halt star formation within infalling galaxies onto a
cluster or group \citep{cowie77a, cowie77b}. Like ram--pressure stripping,
evaporation occurs most efficiently in the central regions of
clusters. However, the gas density and temperature in the outskirts of
clusters or in groups are sufficient to make these processes efficient in
less extreme environs than the centers of the most massive clusters
\citep{fujita04, hester06}.

However, the results of this paper show that environment--dependent (or
environment--related) processes are not responsible for driving the decline
in the cosmic star--formation rate. Instead, our analysis supports the
general picture of a smooth decline in the typical galaxy SFR, with only a
relatively minor impact from environment--related mechanisms such as
mergers, evaporation associated with conductive heating, and ram--pressure
stripping. The star--formation rate in both dense and underdense regions
has decreased rapidly since $z \sim 1$, rather than primarily declining in
group and cluster--like environments.

\begin{figure}[h]
\centering
\plotone{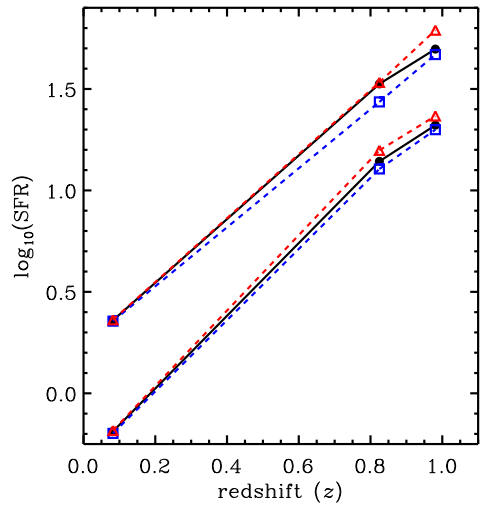}
\caption{The mean and median galaxy SFR as a function of redshift for blue
  galaxies in the SDSS--B and DEEP2--C samples (black filled circles). The
  blue galaxy populations are selected according to Equation
  \ref{willmer_cut}, with an offset of 0.14 magnitudes for the SDSS as
  described in \S \ref{sec_data1}). The DEEP2--C sample is divided into two
  subsamples according to redshift ($0.75 < z < 0.9$ and $0.9 < z <
  1.05$). The blue squares and red triangles show the evolution in the mean
  and median SFR for blue galaxies in the lowest--density (bottom 20\%,
  blue squares) and highest--density (top 20\%, red triangles) environments
  in the corresponding redshift regime. Note that the values tracing the
  evolution in the mean SFR are offset by $+0.25$ in $\log_{10}({\rm SFR})$
  to facilitate display. The overall decline in the mean and median SFR
  with redshift is significantly greater than the dependence of the mean
  and median SFR on environment at either $z \sim 0$ or $z \sim 1$. Note
  that the star--formation rates are given in units of $h^{-2} {\rm
    M}_{\sun} / {\rm yr}$.}
\label{mean_sfr_z}
\end{figure}

Figure \ref{mean_sfr_z} shows the decline in the average (and median) SFR
for the blue galaxy population within the volume--limited [relative to
$M_{B}^{*}(z)$] DEEP2--C and SDSS--B samples. The evolution in the mean SFR
is much stronger than the dependence of mean SFR on environment at $z \sim
1$ or $z \sim 0$ (cf.\ Fig.\ \ref{sfr_environ_fig}); the average (or
median) SFR drops by a factor of $\sim \! 20$ between $z \sim 1$ and $z
\sim 0.1$, while the contrast between the mean SFR in the highest-- and
lowest--density regions is less than a factor $2$ at both $z \sim 1$ and $z
\sim 0.1$. Changing the rate of $M_B^{*}$ evolution used to define these
samples (i.e., the parameter $Q$ in Equation \ref{magcut_eqn}) changes
these results only modestly; the growth in the mean SFR for these matched
samples ranges from a factor of $\sim \! 18$ to $\sim \! 16$ for $Q =
-1.37$ or $Q = -1$. While our results show that environment is clearly
correlated with star--formation activity at $z < 1$ and that there is
significant evolution in the relationship between mean SFR and galaxy
overdensity from $z \sim 1$ to $z \sim 0$, the strength of the SFR--density
relation and its evolution at $z < 1$ is a small perturbation on the
overall decline in the global star--formation rate space density. If
environment--related processes played a dominant role in the cosmic
star--formation history over the last 7 Gyr, then we would expect the
decline of the cosmic star formation rate to be associated with only a
subset of environments.

While galaxy environment does not dictate the evolution of the global
star--formation activity at $z < 1$, it does play a central role in both
the complete quenching of star formation (i.e., the creation of ``red and
dead'' galaxies) and the evolution of the bright, blue galaxy (BBG)
population. As first shown by \citet{cooper06}, all major features of the
color--density relation were in place by $z \sim 1$. This significant
relationship between rest--frame color and environment, which is echoed in
the sSFR--density relation, is a result of star--formation activity being
completely halted in groups much more frequently than in less--dense
environments. Similarly, by analyzing the evolution of the blue galaxy
fraction in both galaxy groups and the field in DEEP2, \citet{gerke07}
showed that the growth in the abundance of galaxies on the red sequence
\citep[e.g.,][]{brown07, faber07} occurred primarily in groups and clusters
of galaxies at $z < 1.3$ \citep[see also][]{cooper07a}.

Environment plays a related role in the evolution of BBGs. As first
discussed by \citet{cooper06}, in order to reconcile the differences in the
luminosity--environment trends at $z \sim 1$ and $z \sim 0$, the BBGs in
dense environments at $z \sim 1$ must have ceased star formation and moved
onto the red sequence by the present epoch. This evolution may play a
significant role in the evolution of the bright end of the blue galaxy
luminosity function, which sees a significant drop off in number density at
bright magnitudes over the last 7 Gyr \citep[e.g.,][]{willmer06, bundy06,
  zucca06}. This quenching of the BBGs in overdense environments will have
only a small impact on the global SFR, as the most massive galaxies
contribute only a small portion of the total SFR density at $z < 1$
\citep{juneau05, zheng07}.

As this paper was being completed, a parallel analysis of the relationship
between star formation and environment at intermediate redshift was
presented by \citet{elbaz07}. Our results, which are based on a
significantly larger sample and probe to lower levels of star--formation
activity, are in relatively good agreement with those of
\citet{elbaz07}. Both analyses find an inversion (or ``reversal'') in the
SFR--density relation at $z \sim 1$ relative to the local
relation. Furthermore, both studies conclude that major mergers cannot be
the solitary physical mechanism responsible for the correlation between SFR
and galaxy overdensity at intermediate redshift. Using {\it HST}/ACS
imaging data, \citet{elbaz07} conclude that the majority of luminous
infrared galaxies in high--density regions at $z \sim 1$ exhibit normal
spiral morphologies --- i.e., they are not disturbed systems showing
signatures of recent mergers \citep[see also][]{melbourne05}.

The \citet{elbaz07} data set, which is based on deep 24$\mu$m {\it
  Spitzer}/MIPS imaging in the GOODS \citep{dickinson03} fields, provides a
reassuring cross--check to our study based on [O {\small II}] line
luminosities. While multiwavelength analysis in the ECDFS, GOODS, and EGS
fields \citep[e.g.,][]{marcillac07, woo07} will further explore the impact
of dust and metallicity on the connection between environment and star
formation at $z < 1$, the agreement between our work and that of
\citet{elbaz07} indicates that extinction effects are not strongly biasing
our results at $z \sim 1$, and that missing low--SFR galaxies is not
dominating their results.

\section{Summary \& Conclusions}
\label{sec_summary}

In this paper, we present a detailed study of the relationship between star
formation and local environment at both $z \sim 0$ and $z \sim 1$ using
galaxy samples drawn from the SDSS and DEEP2 surveys. We estimate the local
overdensity about each galaxy according to the projected $3^{\rm
  rd}$--nearest--neighbor surface density, and measure star--formation
rates using [O {\small II}] $\lambda3727$\AA\ line luminosities, calibrated
to more robust star--formation indicators. Our principal results are as
follows:

\begin{itemize}

\item We find that the relationships between specific star--formation rate
  and environment at $z \sim 0$ and at $z \sim 1$ are very similar, with
  the mean sSFR decreasing in regions of higher galaxy density. We conclude
  that this trend, like the color--density relation at $z < 1$, is driven
  by the quenching of star formation in regions of high galaxy overdensity
  (i.e., galaxy groups and clusters). At both epochs, we find a close
  correlation between sSFR and rest--frame $U-B$ color.

\item In contrast to the local SFR--density relation, we find an inversion
  in the dependence of mean SFR on local overdensity at $z \sim 1$, such
  that the typical SFR increases in higher--density regions, rather than
  decreasing as at $z\sim 0$. At $z\sim 1$, both the highest-- and
  lowest--SFR galaxies are typically found in denser regions than
  intermediate objects, while at $z\sim 0$, the highest--SFR galaxies
  prefer void--like environments. This reflects the fact that there is a
  significant positive correlation between luminosity and overdensity at
  $z\sim 1$ which is weak or absent locally. These trends are associated
  with the existence a population of bright, massive blue galaxies in dense
  regions at $z \sim 1$, as first discovered by \citet{cooper06}, which
  have no local counterparts. This galaxy population is thought to evolve
  into members of the red sequence from $z \sim 1$ to $z \sim 0$.

\item Environmental effects do not  play a dominant role in shaping the cosmic
  star--formation history at $z < 1$. The dependence of the mean galaxy SFR
  on local galaxy density and its evolution from $z \sim 1$ to $z \sim 0$
  is much weaker than the decline in the global SFR space density
  over the last 7 Gyr.

\end{itemize}

%%%%%%%%%%%%%%%%%%%%%%%
%%% Acknowledgments %%%
%%%%%%%%%%%%%%%%%%%%%%%

\acknowledgments This work was supported in part by NSF grants AST--0071048
AST--0071198, AST--0507428, and AST--0507483 as well as Hubble Space
Telescope Archival grant, HST--AR--10947.01. A.L.C.\ acknowledges support
by NASA through Hubble Fellowship grant HST--HF--01182.01--A, awarded by
the Space Telescope Science Institute, which is operated by AURA Inc.\
under NASA contract NAS 5--26555. M.C.C.\ would like to thank Greg Wirth
and all of the Keck Observatory staff for their help in the acquisition of
the Keck/DEIMOS data. M.C.C.\ would also like to thank Michael Blanton and
David Hogg for their assistance in utilizing the NYU--VAGC data products.

We also wish to recognize and acknowledge the highly significant
cultural role and reverence that the summit of Mauna Kea has always
had within the indigenous Hawaiian community. It is a privilege to be
given the opportunity to conduct observations from this mountain.

%%%%%%%%%%%%%%%%%%%%
%%% Bibliography %%%
%%%%%%%%%%%%%%%%%%%%
%\bibliographystyle{apj} 
%\bibliography{apj-jour,environ_sfr_refs}

\end{document}